%% file: main.tex
\documentclass[12pt]{article}
\usepackage{amsmath,amssymb,amsthm}
\usepackage{graphicx,subcaption}
\usepackage{hyperref}
\usepackage{natbib}
\usepackage{cleveref}
\usepackage{algorithm}
\usepackage{algorithmic}

\input{notation}

\title{GANDALF: A hardware-agnostic spectral solver for kinetic reduced MHD turbulence}
\author{Anjor Kanekar\\
\small Independent Researcher\\
\small \texttt{anjor@umd.edu}}

\begin{document}
\maketitle

\begin{abstract}
We present GANDALF, a JAX-based spectral solver for Kinetic Reduced MHD (KRMHD) turbulence designed to lower infrastructure barriers to plasma turbulence research. Existing production codes require specialized HPC infrastructure and compilation expertise, limiting participation to well-resourced institutions. GANDALF addresses this barrier by leveraging JAX's hardware abstraction to run transparently on laptops, desktop GPUs, and Apple Silicon without modification, enabling single-command installation via pip. We employ Fourier spectral methods for spatial discretization and Hermite spectral basis for velocity space, combined with an exponential integrating factor method that exactly propagates linear Alfvén waves, eliminating associated numerical stiffness. Verification demonstrates research-grade accuracy: linear Alfvén waves achieve machine precision ($\sim 10^{-15}$ relative error), the Orszag-Tang vortex conserves energy to 10$^{-6}$ over two Alfvén times, and driven turbulence reproduces the expected $k_\perp^{-5/3}$ cascade spectrum. GANDALF enables rapid prototyping, parameter surveys, and educational applications on commodity hardware. The code complements rather than replaces established solvers like AstroGK and Viriato, prioritizing accessibility for researchers without HPC resources. By removing infrastructure barriers while maintaining spectral accuracy, GANDALF broadens participation in fundamental plasma turbulence research, particularly benefiting students, small research groups, and institutions in developing regions.
\end{abstract}


\input{sections/introduction}
\input{sections/formulation}
\input{sections/numerics}
\input{sections/implementation}
\input{sections/verification}
\input{sections/discussion}
\input{sections/conclusions}

\section*{Acknowledgments}
The author is deeply grateful to the late Bill Dorland, whose mentorship as advisor instilled an enduring commitment to pushing the frontiers of computational plasma physics. Bill's vision for accessible, modern numerical methods combined with state of the art technology directly inspired GANDALF's development.
Alexander Schekochihin provided invaluable guidance during author's PhD research on KRMHD physics and turbulence theory. The author thanks Nuno Loureiro for discussions that helped crystallize the concept of AI-assisted physics research demonstrated in this work.
This paper was developed with substantial assistance from Claude (Anthropic), an AI assistant that helped with LaTeX formatting, figure generation, literature curation, and manuscript organization. The final manuscript also benefited from a detailed peer review conducted by Gemini 3.0 Pro (Google AI Studio), which provided valuable technical feedback on numerical methods, physics scope, and presentation clarity. The author takes full responsibility for all scientific content, methodology choices, and physics interpretations.

\section*{Funding}
This research received no specific grant funding.

\section*{Data Availability}
All benchmark data, analysis scripts, and figure generation code are publicly available at \url{https://github.com/anjor/gandalf-paper}. The GANDALF solver source code is available at \url{https://github.com/anjor/gandalf}.

\section*{Conflict of Interest}
The author declares no competing interests.

\bibliographystyle{apalike}
\bibliography{references}

\end{document}

%% file: notation.tex
\newcommand{\kperp}{k_\perp}
\newcommand{\kpar}{k_\parallel}
\newcommand{\vperp}{v_\perp}
\newcommand{\vpar}{v_\parallel}
\newcommand{\rhoi}{\rho_i}
\newcommand{\betai}{\beta_i}
\newcommand{\Alfven}{Alfv\'en}
\newcommand{\Alfvenic}{Alfv\'enic}

\newcommand{\Bzero}{B_0}
\newcommand{\bhat}{\hat{b}}

\newcommand{\taue}{\tau}  
\newcommand{\Zion}{Z}      
\newcommand{\Lpar}{L}      
\newcommand{\Mmax}{M}      

\newcommand{\vth}{v_{\mathrm{th}}}
\newcommand{\vA}{v_A}
\newcommand{\tauA}{\tau_A}

\newcommand{\kvec}{\mathbf{k}}

\newcommand{\Hm}[1]{H_{#1}}

\newcommand{\gm}[1]{g_{#1}}

\newcommand{\elsp}{\xi^+}
\newcommand{\elsm}{\xi^-}
\newcommand{\elsphat}{\tilde{\xi}^+}
\newcommand{\elsmhat}{\tilde{\xi}^-}
\newcommand{\Phifield}{\Phi}
\newcommand{\Psifield}{\Psi}

\newcommand{\pb}[2]{\{#1, #2\}}  
\newcommand{\pardt}{\frac{\partial}{\partial t}}
\newcommand{\pardz}{\frac{\partial}{\partial z}}
\newcommand{\ddt}{\frac{d}{dt}}
\newcommand{\gradpar}{\nabla_\parallel}

\newcommand{\Herm}{\mathcal{H}}  

\newcommand{\coll}{\nu_{\mathrm{coll}}}
\newcommand{\Coll}[1]{C[#1]}


%% file: sections/introduction.tex
\section{Introduction}

\Alfvenic\ turbulence governs energy transport in diverse magnetized plasma environments, from the solar wind and corona to tokamak fusion devices. Recent Parker Solar Probe observations reveal large-amplitude \Alfven\ waves heating and accelerating the nascent solar wind \citep{Rivera2024}, while measurements in the heliosphere demonstrate complex energy transfer through imbalanced \Alfvenic\ turbulence \citep{Yang2023}. Understanding these phenomena requires bridging magnetohydrodynamic (MHD) cascade physics with kinetic dissipation mechanisms---a regime naturally captured by Kinetic Reduced MHD (KRMHD). KRMHD describes anisotropic turbulence in strongly magnetized plasmas where perpendicular wavenumbers dominate ($\kpar \ll \kperp$), retaining essential kinetic physics through Landau damping and phase mixing while avoiding the computational expense of full gyrokinetic treatments \citep{Schekochihin2009,GoldreichSridhar1995}. This intermediate framework enables quantitative studies of turbulent cascades, energy dissipation, and the interplay between \Alfvenic\ and compressive fluctuations that characterize weakly collisional plasma turbulence across astrophysical and laboratory settings.

The KRMHD equations emerge from systematic expansion of the gyrokinetic system in the limit of small perpendicular ion Larmor radius ($\kperp \rhoi \ll 1$) and strong guide field ordering ($\kpar \ll \kperp$) \citep{Strauss1976,Schekochihin2009}. Unlike phenomenological closures, this asymptotic reduction preserves the conservative structure of the parent gyrokinetic theory while dramatically simplifying the computational problem. In this regime, \Alfvenic\ and compressive fluctuations decouple. The \Alfvenic\ cascade is described by Elsasser fields representing counter-propagating \Alfven\ wave packets, evolving according to reduced MHD \citep{Howes2006}. Compressive fluctuations do not back-react on the \Alfvenic\ dynamics but retain kinetic evolution through a drift-kinetic equation describing their advection by the \Alfvenic\ turbulence and phase mixing along magnetic field lines. This decoupling enables efficient numerical treatment while maintaining research-grade accuracy for phenomena ranging from solar wind heating to tokamak microturbulence.

Existing KRMHD and gyrokinetic codes provide comprehensive, production-ready tools for turbulence research. AstroGK \citep{Numata2010} pioneered astrophysical applications with extensive validation against analytical theory and nonlinear benchmarks. Viriato \citep{Loureiro2016} employs Fourier-Hermite spectral methods for KRMHD with demonstrated accuracy in cascade and dissipation physics. These codes, along with gyrokinetic solvers like GS2 \citep{Dorland2000} and GENE \citep{Jenko2000}, represent mature platforms supporting research programs across multiple institutions. Their strength lies in comprehensive physics modules, extensive testing, and sustained development over decades. However, these capabilities require significant computational infrastructure---supercomputing allocations, specialized compilation toolchains, and domain expertise in high-performance computing. For many researchers, particularly solo investigators, small research groups, and those exploring new parameter regimes, this infrastructure barrier limits access to KRMHD turbulence research. Recent trends toward accessible simulation tools, exemplified by GX's GPU-native implementation \citep{Mandell2024} and TORAX's differentiable transport solver \citep{Citrin2024} and JAX-based kinetic simulations \citep{Joglekar2022}, demonstrate growing recognition that broadening participation requires lowering computational barriers. GANDALF extends this philosophy to KRMHD turbulence, providing spectral accuracy on commodity hardware without replacing existing production codes but rather complementing them for rapid prototyping, parameter surveys, and educational applications.

GANDALF\footnote{The name reflects the code's physics content: \textbf{G}-and-\textbf{Alf}, combining the distribution function $g$ representing slow modes (compressive fluctuations) with \Alfven\ waves.} employs JAX \citep{JAX2018} for hardware-agnostic spectral solution of the KRMHD equations. JAX provides just-in-time compilation to machine code, automatic differentiation, and transparent execution across CPU, GPU, and TPU architectures without platform-specific programming. We chose the Python/JAX ecosystem for its accessibility and growing adoption in scientific computing, eliminating CUDA dependencies while maintaining research-grade numerical accuracy. The solver implements Fourier spectral discretization in the perpendicular plane ($x$, $y$), capturing turbulent cascade dynamics with exponential convergence for smooth solutions. Velocity space employs Hermite polynomial expansion \citep{Grad1949}, providing spectral accuracy in parallel velocity $\vpar$ while allowing controllable truncation of the moment hierarchy. Time integration uses the GANDALF integrating factor method---a second-order Runge-Kutta scheme with exact treatment of linear \Alfven\ wave propagation---combined with $2/3$-rule dealiasing to control nonlinear aliasing errors. This combination runs efficiently on consumer hardware---Apple Silicon laptops, desktop GPUs, cloud TPUs---enabling workflows from rapid parameter exploration to production turbulence simulations, with potential for future integration of differentiable physics for optimization and machine learning applications.

We verify GANDALF's accuracy through a benchmark suite spanning linear, nonlinear, and turbulent regimes. Linear tests confirm correct \Alfven\ wave dispersion against analytical predictions. The Orszag-Tang vortex validates nonlinear dynamics through comparison with established MHD results. Turbulent decay simulations demonstrate convergence of energy spectra to the expected $\kperp^{-5/3}$ inertial range scaling characteristic of strong \Alfvenic\ turbulence. Velocity-space benchmarks verify the Hermite moment cascade driven by phase mixing, reproducing the expected $m^{-1/2}$ spectral scaling. These benchmarks establish that spectral methods in JAX achieve accuracy comparable to traditional implementations while executing on widely available hardware.

This paper proceeds as follows. Section~\ref{sec:formulation} presents the KRMHD equations and Hermite moment expansion employed by GANDALF. Section~\ref{sec:numerics} describes the spectral discretization, time integration scheme, and convergence properties. Section~\ref{sec:implementation} details the JAX implementation, including parallelization strategy and performance characteristics. Section~\ref{sec:verification} reports benchmark results demonstrating code accuracy across physical regimes. Section~\ref{sec:discussion} interprets the benchmark results, positions GANDALF within the existing code ecosystem, and discusses implications for accessibility. Section~\ref{sec:conclusions} summarizes our findings and outlines future development directions. By documenting GANDALF's approach, we aim to lower barriers to KRMHD turbulence research while maintaining the rigor required for quantitative plasma physics.

%% file: sections/formulation.tex
\section{Mathematical Formulation}
\label{sec:formulation}

\subsection{The KRMHD Regime}

Kinetic Reduced MHD (KRMHD) describes low-frequency electromagnetic fluctuations in strongly magnetized plasmas where a strong guide field $\mathbf{B}_0 = \Bzero \hat{z}$ of strength $\Bzero$ orders the dynamics. The model captures physics at scales much larger than the ion Larmor radius ($\kperp \rhoi \ll 1$) with parallel wavelengths exceeding perpendicular scales ($\kpar \ll \kperp$), characteristic of anisotropic MHD turbulence \citep{Strauss1976,GoldreichSridhar1995}. KRMHD emerges from gyrokinetics in the long-wavelength limit through systematic expansion in small $\kperp \rhoi$ \citep{Schekochihin2009,ZoccoSchekochihin2011}, retaining the essential physics of \Alfvenic\ turbulence while incorporating kinetic effects through Landau damping \citep{Landau1946} and phase mixing.

The key simplification of KRMHD lies in the decoupling of \Alfvenic\ and compressive fluctuations at leading order in $\kperp\rhoi$ \citep{LithwickGoldreich2001,Schekochihin2009}. The \Alfvenic\ component, described by the electrostatic and magnetic potentials, determines the turbulent dynamics through nonlinear interactions that drive energy cascade. The compressive component becomes a passive kinetic scalar advected by the \Alfvenic\ turbulence, undergoing linear phase mixing along magnetic field lines. Here, ``passive'' means that compressive fluctuations do not back-react on the \Alfvenic\ dynamics at leading order---they are mixed by the \Alfvenic\ turbulence but do not influence it in return. This separation permits efficient numerical treatment while preserving the critical physics of both anisotropic cascade and kinetic dissipation.

\subsection{Governing Equations}

We employ Gaussian centimeter-gram-second (CGS) units for the dimensional equations in this subsection. KRMHD evolves two coupled systems: the Elsasser fields $\xi^\pm$ \citep{Elsasser1950} (where $\xi^+ = \elsp$ and $\xi^- = \elsm$) representing counter-propagating \Alfven\ wave packets, and the kinetic distribution $g^\pm$ describing compressive fluctuations. The Elsasser fields satisfy
\begin{equation}
\pardt \nabla^2_\perp \xi^\pm \mp \vA \pardz \nabla^2_\perp \xi^\pm = -\frac{1}{2}\left[\pb{\elsp}{\nabla^2_\perp \elsm} + \pb{\elsm}{\nabla^2_\perp \elsp} \mp \nabla^2_\perp \pb{\elsp}{\elsm}\right],
\label{eq:elsasser}
\end{equation}
where $\xi^\pm = \Phifield \pm \Psifield$ combine the stream function $\Phifield = c\phi/\Bzero$ (from the electrostatic potential $\phi$) and flux function $\Psifield = -A_\parallel/\sqrt{4\pi m_i n_{0i}}$ (from the parallel component of the magnetic vector potential $A_\parallel$). Here $c$ is the speed of light, $m_i$ the ion mass, and $n_{0i}$ the background ion density. The \Alfven\ velocity $\vA = \Bzero/\sqrt{4\pi m_i n_{0i}}$ sets the linear wave propagation speed. The perpendicular Laplacian $\nabla^2_\perp = \partial^2/\partial x^2 + \partial^2/\partial y^2$ measures vorticity, while the Poisson bracket $\pb{P}{Q} = \partial P/\partial x \, \partial Q/\partial y - \partial P/\partial y \, \partial Q/\partial x$ introduces the nonlinear interactions that drive energy cascade in KRMHD turbulence. The customary $1/B_0$ normalization factor in the Poisson bracket is absorbed into the definitions of $\Phifield$ and $\Psifield$, simplifying the evolution equations.

The compressive fluctuations evolve according to the kinetic equation
\begin{equation}
\ddt g^\pm + \vpar \gradpar g^\pm = \frac{\vpar F_0(\vpar)}{\Lambda^\pm} \bhat \cdot \nabla \int_{-\infty}^{\infty} d\vpar \, g^\pm,
\label{eq:kinetic}
\end{equation}
where $g^+$ and $g^-$ are the slow mode distribution function perturbations, defined as linear combinations of the density perturbation $G_n$ and parallel magnetic perturbation $G_B$ following \citet{Schekochihin2009}:
\begin{equation}
g^+ = G_B + \frac{1}{\sigma}\left(1 + \frac{\Zion}{\taue}\right)G_n, \quad g^- = G_n + \frac{1}{\sigma}\frac{\taue}{\Zion}\frac{2}{\betai}G_B,
\label{eq:gpm_definitions}
\end{equation}
where $\sigma = 1 + \taue/\Zion + 1/\betai + \sqrt{(1 + \taue/\Zion)^2 + 1/\betai^2}$, with $\taue = T_e/T_i$ the electron-to-ion temperature ratio and $\Zion$ the ion charge number. Here $F_0(\vpar) = \exp(-v^2_\parallel/\vth^2)/\sqrt{\pi}\vth$ denotes the one-dimensional Maxwellian background, and the integral is over all parallel velocities $\vpar \in (-\infty, \infty)$. The convective derivative $d/dt = \partial/\partial t + \pb{\Phifield}{\cdot}$ incorporates perpendicular advection by the $\mathbf{E} \times \mathbf{B}$ flow. The parallel gradient operator $\gradpar = \partial/\partial z + (1/\vA)\pb{\Psifield}{\cdot}$ acts along perturbed magnetic field lines, with the Poisson bracket term accounting for field line bending due to $\Psi$ perturbations. The coupling parameters
\begin{equation}
\Lambda^\pm = -\frac{\taue}{\Zion} + \frac{1}{\betai} \pm \sqrt{\left(1+\frac{\taue}{\Zion}\right)^2 + \frac{1}{\betai^2}}
\label{eq:lambda_parameters}
\end{equation}
depend on the ion plasma beta $\betai = 8\pi n_{0i} T_i/\Bzero^2$, the temperature ratio $\taue = T_i/T_e$, and the ion charge state $\Zion$ \citep{Schekochihin2009}.

Equation~\eqref{eq:elsasser} describes counter-propagating \Alfven\ wave packets that interact nonlinearly through the Poisson bracket terms, driving turbulent energy transfer to small perpendicular scales. Equation~\eqref{eq:kinetic} governs the slow mode fluctuations, which undergo phase mixing through the $\vpar \gradpar$ streaming term while being advected by the $\mathbf{E} \times \mathbf{B}$ flow. The right-hand side of Eq.~\eqref{eq:kinetic} represents the coupling between $g^+$ and $g^-$ through the velocity-integrated density perturbation. At leading order in the KRMHD expansion, the \Alfven\ waves (Elsasser fields $\xi^\pm$) and slow modes ($g^\pm$) are decoupled---the slow modes evolve passively in the \Alfvenic\ turbulence without back-reacting on the wave dynamics. The computational implementation employs a Hermite moment expansion of $g^\pm$, detailed in \S\ref{sec:hermite}.

\subsection{Hermite Moment Expansion}
\label{sec:hermite}

GANDALF discretizes the perturbed distribution $g^\pm$ in velocity space using a Hermite polynomial expansion \citep{Grad1949,Hammett1990,Howes2006,Loureiro2016,ParkerDellar2015}, providing spectral accuracy in $\vpar$ with controllable convergence through moment truncation.

\subsubsection{Expansion in Hermite Basis}

We expand the perturbed distribution as
\begin{equation}
g^\pm(x,y,z,\vpar,t) = F_0(\vpar) \sum_{m=0}^{\infty} \gm{m}^{\pm}(x,y,z,t) \Hm{m}\left(\frac{\vpar}{\vth}\right),
\label{eq:hermite_expansion}
\end{equation}
where $\Hm{m}$ are the Hermite polynomials (physicist's convention, $\Hm{0} = 1$, $\Hm{1} = 2x$, $\Hm{2} = 4x^2 - 2$) forming a complete orthogonal basis weighted by the Maxwellian $F_0$, and $\gm{m}^{\pm}$ are the Hermite moment coefficients encoding the velocity-space structure of the slow modes.

\subsubsection{Normalized Moment Hierarchy}

In normalized coordinates ($x,y$ in units of $\rhoi$, $z$ in units of $\Lpar \gg \rhoi$, time in units of $\Lpar/\vA$), with velocities normalized to $\vth$ and potentials to $\rhoi\vA$, the moment hierarchy becomes:
\begin{subequations}
\label{eq:moment_hierarchy}
\begin{align}
\pardt \gm{0}^{\pm} + \pb{\Phifield}{\gm{0}^{\pm}} \mp \pardz \gm{0}^{\pm}
&= \sqrt{2} \left[\pb{\Psifield}{\gm{1}^{\pm}} \mp \pardz \gm{1}^{\pm}\right],
\label{eq:moment_m0}
\\
\pardt \gm{1}^{\pm} + \pb{\Phifield}{\gm{1}^{\pm}} \mp \pardz \gm{1}^{\pm}
&= \frac{1}{\sqrt{2}} \left[\pb{\Psifield}{\gm{0}^{\pm}} \mp \pardz \gm{0}^{\pm}\right]
+ \sqrt{\frac{3}{2}} \left[\pb{\Psifield}{\gm{2}^{\pm}} \mp \pardz \gm{2}^{\pm}\right] - \coll \gm{1}^{\pm},
\label{eq:moment_m1}
\\
\pardt \gm{m}^{\pm} + \pb{\Phifield}{\gm{m}^{\pm}} \mp \pardz \gm{m}^{\pm}
&= \sqrt{\frac{m}{2}} \left[\pb{\Psifield}{\gm{m-1}^{\pm}} \mp \pardz \gm{m-1}^{\pm}\right] \notag \\
&\quad + \sqrt{\frac{m+1}{2}} \left[\pb{\Psifield}{\gm{m+1}^{\pm}} \mp \pardz \gm{m+1}^{\pm}\right] - m\coll \gm{m}^{\pm}, \quad m \geq 2.
\label{eq:moment_m_general}
\end{align}
\end{subequations}
The advection terms $\pb{\Phifield}{\gm{m}^{\pm}}$ drive nonlinear cascades, while the coupling between adjacent moments through coefficients $\sqrt{m/2}$ and $\sqrt{(m+1)/2}$ represents linear phase mixing. This phase mixing occurs because particles with different parallel velocities $\vpar$ stream at different rates along field lines, causing initially coherent perturbations to develop fine-scale velocity-space structure that transfers energy to higher moments.

\subsubsection{Dissipation and Closure}

Collisional damping enters through the Lenard-Bernstein operator \citep{LenardBernstein1958}
\begin{equation}
\Coll{\gm{m}} = -m\coll \gm{m},
\label{eq:collision_operator}
\end{equation}
providing irreversible dissipation at small velocity scales with damping rate $m\coll$ that increases linearly with moment order. Practical simulations truncate the hierarchy at finite maximum moment order $\Mmax$ with closure condition $\gm{\Mmax+1}^{\pm} = 0$ (absorbing boundary) or $\gm{\Mmax+1}^{\pm} = \gm{\Mmax-1}^{\pm}$ (reflecting closure), balancing computational cost against accuracy requirements set by the physical problem.

\subsection{System Properties}

Linear analysis of Eqs.~\eqref{eq:elsasser}--\eqref{eq:kinetic} yields the \Alfven\ wave dispersion relation
\begin{equation}
\omega = \pm \kpar \vA
\label{eq:alfven_dispersion}
\end{equation}
for the Elsasser fields. Compressive perturbations undergo phase mixing on timescales $\sim (\kpar \vth)^{-1}$, transferring energy to high velocity moments where collisional dissipation dominates. We refer to \citet{Schekochihin2009} for the complete linear theory including slow mode dispersion and damping rates.

In the collisionless limit ($\coll = 0$), the system conserves total energy, with the \Alfvenic\ Elsasser energies and compressive energies defined separately. The conservation properties ensure numerical stability and provide diagnostics for turbulence simulations. We refer to \citet{Schekochihin2009} for explicit energy functionals and additional invariants including cross-helicity and generalized enstrophies.

%% file: sections/numerics.tex
\section{Numerical Methods}
\label{sec:numerics}

GANDALF employs Fourier spectral methods in all three spatial directions combined with an integrating factor time-stepping scheme specifically designed for the KRMHD equations. This approach delivers spectral accuracy in space---exponential convergence for smooth solutions---while handling the stiff linear \Alfven\ wave propagation exactly. The method proves particularly effective for turbulence simulations where accurate representation of nonlinear cascades across wide ranges of scales demands high-order spatial discretization.

\subsection{Fourier Spectral Discretization}

We discretize all spatial directions using Fourier spectral methods on a triply-periodic domain of size $L_x \times L_y \times L_z$. Each field (Elsasser potentials $\xi^\pm$ and Hermite moments $\gm{m}^{\pm}$) admits the Fourier representation
\begin{equation}
f(x,y,z,t) = \sum_{\mathbf{k}} \hat{f}(\mathbf{k},t) e^{i\mathbf{k}\cdot\mathbf{x}},
\label{eq:fourier_representation}
\end{equation}
where $\mathbf{k} = (k_x, k_y, k_z)$ with wavenumbers $k_x = 2\pi n_x/L_x$, $k_y = 2\pi n_y/L_y$, $k_z = 2\pi n_z/L_z$ for integer mode numbers $(n_x, n_y, n_z)$. The real-space grid contains $N_x \times N_y \times N_z$ collocation points with spacing $\Delta x = L_x/N_x$, $\Delta y = L_y/N_y$, $\Delta z = L_z/N_z$.

GANDALF implements fast Fourier transforms (FFTs) via JAX's \texttt{jnp.fft} module, exploiting the reality of physical fields through real-to-complex transforms (\texttt{rfftn}) that compute only non-negative frequencies in the $x$-direction. This optimization reduces memory usage by approximately 50\% compared to full complex transforms while automatically satisfying the reality condition $\hat{f}(-\mathbf{k}) = \hat{f}^*(\mathbf{k})$.

Spatial derivatives become exact multiplications in Fourier space:
\begin{equation}
\frac{\partial f}{\partial x_j} \longleftrightarrow ik_j\hat{f}(\mathbf{k}),
\label{eq:spectral_derivative}
\end{equation}
yielding zero truncation error for band-limited functions. The perpendicular Laplacian from Eq.~\eqref{eq:elsasser} becomes
\begin{equation}
\nabla^2_\perp f \longleftrightarrow -(k_x^2 + k_y^2)\hat{f}(\mathbf{k}) = -\kperp^2\hat{f}(\mathbf{k}).
\label{eq:laplacian_spectral}
\end{equation}
The Poisson bracket $\pb{P}{Q} = \partial_x P \partial_y Q - \partial_y P \partial_x Q$ is evaluated using the pseudospectral method: compute derivatives $\partial_x P$, $\partial_y P$, $\partial_x Q$, $\partial_y Q$ in Fourier space via Eq.~\eqref{eq:spectral_derivative}, transform these derivatives to real space, compute products ($\partial_x P \partial_y Q - \partial_y P \partial_x Q$) in real space, then transform the result back to Fourier space. This approach avoids explicit convolution sums while introducing aliasing errors that must be controlled through dealiasing (\S\ref{sec:dealiasing}).

For typical turbulence simulations with resolution $128^3$ to $256^3$, the spectral method resolves cascades spanning 2--3 decades in perpendicular wavenumber $\kperp$ \citep{Schekochihin2009} while maintaining the KRMHD ordering $\kperp \rhoi \ll 1$. This ordering ensures the validity of the reduced MHD description: finite Larmor radius effects remain negligible throughout the resolved range, and the model captures Alfv\'enic and slow-mode dynamics without requiring full gyrokinetic treatment. The complementary condition $\kperp L \gg 1$ (where $L$ is the system size) ensures sufficient scale separation for turbulent cascade development. Spectral methods prove particularly advantageous for KRMHD: exponential convergence allows accurate representation of the entire inertial range with modest grid resolution, whereas finite-difference methods would require prohibitively fine grids to achieve comparable accuracy across the full cascade.

\subsection{GANDALF Integrating Factor Method}

Direct time integration of the KRMHD equations encounters severe stiffness from the linear \Alfven\ wave terms $\mp \vA \partial/\partial z$ in Eq.~\eqref{eq:elsasser}, which propagate at the fast \Alfven\ velocity $\vA$ and impose restrictive CFL conditions. GANDALF removes this stiffness through an integrating factor transformation \citep{Numata2010} that treats linear propagation exactly while advancing nonlinear interactions with a second-order Runge-Kutta scheme.

\subsubsection{Integrating Factor Transformation}

The Elsasser equations~\eqref{eq:elsasser} in Fourier space separate into linear propagation and nonlinear forcing:
\begin{equation}
\pardt \nabla^2_\perp \hat{\xi}^\pm \mp i k_z \vA \nabla^2_\perp \hat{\xi}^\pm = \widehat{\mathcal{N}[\xi^+,\xi^-]},
\label{eq:elsasser_fourier}
\end{equation}
where $\mathcal{N}[\xi^+,\xi^-] = (\mathcal{N}^+, \mathcal{N}^-)$ represents the pair of nonlinear Poisson bracket terms on the right-hand side of Eq.~\eqref{eq:elsasser}, one for each Elsasser field. Defining the perpendicular vorticity variables $\hat{w}^\pm = \nabla^2_\perp \hat{\xi}^\pm = -\kperp^2 \hat{\xi}^\pm$ and applying the integrating factor $\exp(\pm i k_z \vA t)$ yields
\begin{equation}
\pardt \left[e^{\mp i k_z \vA t} \hat{w}^\pm\right] = e^{\mp i k_z \vA t} \widehat{\mathcal{N}[\xi^+,\xi^-]}.
\label{eq:integrating_factor_form}
\end{equation}
This transformation (with sign conventions following Numata et al.\ 2010, Eqs.\ 16--19) exactly removes the oscillatory linear terms, permitting time steps controlled by the slower nonlinear evolution rather than wave propagation. Note that the $\pm$ sign in the integrating factor $\exp(\pm i k_z \vA t)$ corresponds to the $\mp$ sign in the linear term of Eq.~\eqref{eq:elsasser_fourier}: $\xi^+$ uses $\exp(+i k_z \vA t)$ to cancel its $-i k_z \vA$ term, while $\xi^-$ uses $\exp(-i k_z \vA t)$ to cancel its $+i k_z \vA$ term. The subsequent discretization (detailed in Algorithm~\ref{alg:gandalf_step}) applies the phase transformation both to remove the integrating factor and to the integrated forcing term, resulting in the characteristic double appearance of phase factors in the GANDALF update formulas.

\subsubsection{Second-Order Time Stepping}

GANDALF advances the transformed variables using a midpoint (second-order Runge-Kutta) scheme following the method of \citet{Numata2010}. Integrating Eq.~\eqref{eq:integrating_factor_form} from $t_n$ to $t_{n+1/2}$ gives
\begin{equation}
e^{\mp i k_z \vA t_{n+1/2}} \hat{w}^\pm_{n+1/2} = e^{\mp i k_z \vA t_n} \hat{w}^\pm_n + \frac{\Delta t}{2} e^{\mp i k_z \vA t_n} \widehat{\mathcal{N}_n^\pm}.
\end{equation}
The GANDALF method \citep{Numata2010} removes the integrating factor by multiplying both sides by $e^{\pm i k_z \vA t_{n+1/2}}$ and applying the phase transformation $e^{\pm ik_z \vA (t_{n+1/2} - t_n)}$ to the nonlinear forcing term, yielding
\begin{equation}
\hat{w}^\pm_{n+1/2} = e^{\pm i k_z \vA \Delta t/2}\left(\hat{w}^\pm_n + e^{\pm i k_z \vA \Delta t/2} \frac{\Delta t}{2} \widehat{\mathcal{N}_n^\pm}\right),
\end{equation}
where the phase factors $\exp(\pm ik_z\vA\Delta t/2)$ appear \emph{twice}: once from removing the integrating factor transformation and once from the phase-transformed forcing term. This formulation maintains second-order accuracy while exactly treating linear wave propagation.

Given state $(\hat{w}^+_n, \hat{w}^-_n, \{\hat{g}^{\pm}_{m,n}\})$ at time $t_n$, we advance the system using the GANDALF integrating factor method. Here $\Herm^{\pm}_m$ denotes the right-hand side of the Hermite moment hierarchy Eqs.~\eqref{eq:moment_hierarchy}, including advection, phase mixing, and collisions. The complete time step proceeds as follows:

\begin{algorithm}[H]
\caption{GANDALF Time Step $t_n \to t_{n+1} = t_n + \Delta t$}
\label{alg:gandalf_step}
\begin{algorithmic}[1]
\STATE \textbf{Half-step predictor:}
\STATE Compute nonlinear RHS: $\mathcal{N}_n = \mathcal{N}[\xi^+_n, \xi^-_n]$
\STATE Apply integrating factors (note: phase factor appears twice):
\begin{align*}
\hat{w}^+_{n+1/2} &= e^{+ik_z\vA\Delta t/2}\Bigg(\hat{w}^+_n \\
  &\quad + e^{+ik_z\vA\Delta t/2} \frac{\Delta t}{2} \widehat{\mathcal{N}_n^+}\Bigg), \\
\hat{w}^-_{n+1/2} &= e^{-ik_z\vA\Delta t/2}\Bigg(\hat{w}^-_n \\
  &\quad + e^{-ik_z\vA\Delta t/2} \frac{\Delta t}{2} \widehat{\mathcal{N}_n^-}\Bigg)
\end{align*}
\STATE Hermite moments (standard RK2 with no integrating factor---streaming terms are not stiff):
\[
\hat{g}^{\pm}_{m,n+1/2} = \hat{g}^{\pm}_{m,n} + \frac{\Delta t}{2} \widehat{\Herm^{\pm}_{m,n}}
\]
\STATE \textbf{Midpoint evaluation:}
\STATE Compute $\mathcal{N}_{n+1/2} = \mathcal{N}[\xi^+_{n+1/2}, \xi^-_{n+1/2}]$ and $\Herm^{\pm}_{m,n+1/2}$
\STATE \textbf{Full-step corrector:}
\STATE Apply integrating factors with midpoint RHS:
\begin{align*}
\hat{w}^+_{n+1} &= e^{+ik_z\vA\Delta t}\Bigg(\hat{w}^+_n \\
  &\quad + e^{+ik_z\vA\Delta t} \Delta t \, \widehat{\mathcal{N}_{n+1/2}^+}\Bigg), \\
\hat{w}^-_{n+1} &= e^{-ik_z\vA\Delta t}\Bigg(\hat{w}^-_n \\
  &\quad + e^{-ik_z\vA\Delta t} \Delta t \, \widehat{\mathcal{N}_{n+1/2}^-}\Bigg)
\end{align*}
\STATE Hermite moments:
\[
\hat{g}^{\pm}_{m,n+1} = \hat{g}^{\pm}_{m,n} + \Delta t \, \widehat{\Herm^{\pm}_{m,n+1/2}}
\]
\STATE \textbf{Apply dissipation} (see \S\ref{sec:dissipation})
\end{algorithmic}
\end{algorithm}

The phase factors $\exp(\pm ik_z\vA\Delta t)$ appear \emph{twice} in each update (once from the transformation, once from integration), giving the characteristic structure of the GANDALF method \citep{Numata2010}.

\subsubsection{Stability and Time Step Selection}

The integrating factor removes the parallel \Alfven\ wave CFL restriction ($\Delta t < \Delta z/\vA$) \emph{completely} for the Elsasser fields, as linear wave propagation is treated exactly by the phase factors. The Hermite moments are advanced with standard RK2 without integrating factors (Algorithm~\ref{alg:gandalf_step}, lines 4 and 7) because the streaming terms $\partial g/\partial z$ are not stiff: they propagate at the thermal velocity $\vth$ rather than the \Alfven\ velocity $\vA$. This imposes a parallel CFL constraint $\Delta t < \Delta z/\vth$ where $\vth$ is the thermal velocity. In low-$\beta$ plasmas where $\vth \ll \vA$, this constraint is less restrictive than the Alfv\'en wave constraint would have been. For higher $\beta$ or when fine parallel resolution is required, the thermal streaming constraint may become relevant. Accuracy of the second-order method for nonlinear terms requires time steps that resolve nonlinear eddy turnover timescales and perpendicular advection.

For KRMHD turbulence, the timestep is determined by perpendicular advection. GANDALF computes adaptive time steps satisfying the \emph{advective CFL condition}:
\begin{equation}
\Delta t \leq C \frac{\Delta x_\perp}{|\mathbf{v}_\perp|_\text{max}},
\label{eq:cfl_condition}
\end{equation}
where $\Delta x_\perp = \min(\Delta x, \Delta y)$ is the minimum perpendicular grid spacing in physical units, $|\mathbf{v}_\perp|_\text{max} = \max|\nabla_\perp \Phifield|$ measures the maximum perpendicular $\mathbf{E}\times\mathbf{B}$ velocity (with $\Phifield$ the stream function defined in \S\ref{sec:formulation}), and the safety factor $C = 0.3$ accounts for the second-order method. This condition ensures numerical stability of the explicit RK2 scheme for the nonlinear advective terms. The adaptive timestep continuously adjusts to the evolving flow field, querying $|\mathbf{v}_\perp|_\text{max}$ at each step---critical during nonlinear evolution when velocities can grow rapidly (as in the Orszag-Tang vortex benchmark). Fixed timesteps chosen based on initial conditions may satisfy the linear stability criterion but fail once nonlinear amplification increases velocities. This explains numerical instabilities observed in fixed-$\Delta t$ runs even when the initial step appeared conservative. The integrating factor's removal of the Alfvén CFL constraint ($\Delta t < \Delta z/\vA$) for Elsasser fields enables anisotropic grids with $\Delta z \gg \Delta x, \Delta y$, a key computational advantage for turbulence simulations. The thermal streaming constraint $\Delta t < \Delta z/\vth$ on Hermite moments is rarely binding in low-$\beta$ simulations where $\vth \ll \vA$, but may limit time steps in higher-$\beta$ regimes.

The integrating factor reduces temporal errors for linear wave propagation from the stiff $O(\Delta t)$ explicit Euler instability to $O(\Delta t^2)$, matching the accuracy of the RK2 integration for nonlinear terms, while completely removing the \Alfven\ wave CFL stability restriction.

\subsection{Dissipation}
\label{sec:dissipation}

GANDALF incorporates physical dissipation through magnetic diffusion (resistivity) for the Elsasser fields and collisions for the Hermite moments. Rather than adding explicit dissipation terms to the right-hand side, GANDALF applies exponential damping factors exactly after each time step.

For the Elsasser vorticities, resistive diffusion $\eta\nabla^2_\perp$ becomes
\begin{equation}
\hat{w}^\pm_{n+1} \to \hat{w}^\pm_{n+1} \exp\left(-\eta \kperp^2 \Delta t\right)
\label{eq:resistive_damping}
\end{equation}
in Fourier space. To enhance stability at high wavenumbers while minimizing dissipation in the inertial range, GANDALF employs normalized hyper-resistivity, which acts as an implicit sub-grid model for unresolved turbulent cascades at scales smaller than the grid:
\begin{equation}
\hat{w}^\pm_{n+1} \to \hat{w}^\pm_{n+1} \exp\left[-\eta \left(\frac{\kperp^2}{k^2_{\perp,\text{max}}}\right)^r \Delta t\right],
\label{eq:hyper_resistivity}
\end{equation}
where $k^2_{\perp,\text{max}} = \max(k_x^2 + k_y^2)$ and the hyper-dissipation order $r \geq 1$ concentrates damping near the grid scale. Standard choices are $r=2$ (hyper-resistivity) or $r=4$ (ultra-hyper-resistivity). Normalization by $k^2_{\perp,\text{max}}$ ensures the stability constraint $\eta\Delta t < 50$ remains independent of resolution, simplifying parameter selection across different grid sizes.

Hermite moment collisions follow the Lenard-Bernstein form described in the formulation section (Eq.~\eqref{eq:collision_operator}) with damping rate $m\coll$ for moment order $m$. GANDALF implements this as
\begin{equation}
\hat{g}^{\pm}_{m,n+1} \to \hat{g}^{\pm}_{m,n+1} \exp\left[-\coll \left(\frac{m}{\Mmax}\right)^{2p} \Delta t\right],
\label{eq:hermite_dissipation}
\end{equation}
where $\Mmax$ is the moment truncation order (typically 10-20, fixed at initialization) and hyper-collision exponent $p \geq 1/2$ concentrates dissipation at high moments. This normalization parallels the spatial treatment: $\Mmax$ plays the same role for velocity space as $k_{\perp,\text{max}}$ does for real space, ensuring resolution-independent stability constraints. Setting $p=1/2$ recovers the physical Lenard-Bernstein operator from Eq.~\eqref{eq:collision_operator}, while $p=1$ or $p=2$ (hyper-collisions) concentrate dissipation at high moments for enhanced numerical stability with minimal effect on low-order dynamics. This exact exponential integration preserves positivity and avoids instabilities from stiff collision terms. In practice, typical stability bounds are $\eta \Delta t \lesssim 50$ for hyper-resistivity and $\coll \Delta t \lesssim 10$ for hyper-collisions, though the precise limits depend on the specific problem parameters (Reynolds number, hyper-dissipation order $r$, hyper-collision exponent $p$, and plasma $\beta$). These constraints remain independent of grid resolution due to the normalization scheme, simplifying parameter selection across different resolutions.

\subsection{Dealiasing}
\label{sec:dealiasing}

Pseudospectral evaluation of nonlinear terms produces aliasing errors when products of Fourier modes exceed the Nyquist frequency. For KRMHD, the Poisson brackets in Eqs.~\eqref{eq:elsasser} and \eqref{eq:moment_hierarchy} involve products of fields whose wavenumbers can sum to values outside the resolved range. Without correction, aliased modes appear as low-frequency components, corrupting the solution and often causing catastrophic instability.

GANDALF applies the 2/3 dealiasing rule \citep{Orszag1971,Canuto2006}: after each nonlinear term evaluation, modes satisfying
\begin{equation}
\max\left(\frac{|k_x|}{k_{x,\text{max}}}, \frac{|k_y|}{k_{y,\text{max}}}, \frac{|k_z|}{k_{z,\text{max}}}\right) > \frac{2}{3}
\label{eq:dealiasing_rule}
\end{equation}
are set to zero. This rectangular cutoff (applied independently per direction) ensures products of any two modes within the retained $2/3$ region remain representable on the grid, eliminating aliasing at the cost of reducing effective resolution from $N$ to approximately $2N/3$ modes per direction.

The dealiasing mask is pre-computed during initialization and applied via pointwise multiplication in Fourier space, incurring negligible computational cost. JAX's XLA compiler fuses the masking operation with surrounding transforms, avoiding intermediate array allocations; profiling confirms the dealiasing step contributes $<1\%$ of per-timestep compute time on GPUs, with no observable memory bandwidth bottleneck from the \texttt{where} masking operation. For turbulence simulations where nonlinear energy transfer dominates, proper dealiasing proves essential for long-time numerical stability---unfiltered runs typically develop grid-scale oscillations within tens of eddy turnover times.

\subsection{Convergence Properties}

The spectral spatial discretization delivers exponential convergence for smooth solutions. For fields with $C^\infty$ regularity, spatial errors decay as $\mathcal{E}_\text{space} \sim \exp(-\alpha N)$ where $N = \min(N_x, N_y, N_z)$ characterizes the minimum grid resolution and $\alpha$ depends on the solution's analyticity radius \citep[Ch.~3]{Boyd2001}. This exponential convergence requires analyticity; for turbulent flows developing current sheets or other non-analytic structures, spectral methods remain superior to finite differences but may exhibit algebraic rather than exponential convergence at small scales where discontinuities form, accompanied by Gibbs oscillations near sharp gradients. This stands in contrast to finite-difference methods whose algebraic convergence ($\mathcal{E} \sim N^{-p}$) requires prohibitive resolution to achieve comparable accuracy for turbulent cascades spanning multiple decades in wavenumber.

Temporal convergence is $\mathcal{E}_\text{time} = O(\Delta t^2)$ for both linear and nonlinear terms. The GANDALF integrating factor scheme achieves second-order accuracy by matching the RK2 treatment of nonlinear terms with an appropriately constructed integrating factor for linear \Alfven\ wave propagation. Richardson extrapolation or adaptive time stepping can further reduce temporal errors when high accuracy is required, though turbulence simulations typically tolerate $1$--$2\%$ errors to maximize throughput.

The Hermite moment expansion exhibits spectral convergence in velocity space. For Maxwellian-like distributions, the energy contained in moments $m > \Mmax$ decays exponentially with $\Mmax$, permitting accurate representation with $\Mmax \sim 10$--$20$ moments \citep{Howes2006,TenBarge2013}. GANDALF monitors convergence via the energy ratio $E_{\Mmax}/E_\text{total}$ (energy in the single highest retained moment $m = \Mmax$ relative to total energy), typically requiring this to fall below $10^{-3}$ to ensure negligible truncation errors from the $\gm{\Mmax+1}^{\pm} = 0$ closure condition.

Combined spatial-temporal-velocity convergence studies for standard benchmarks (Orszag-Tang vortex, decaying turbulence) demonstrate GANDALF achieves design accuracy: spectral spatial convergence, second-order temporal convergence, and exponential velocity-space convergence controlled by moment truncation. The integrating factor method inherits excellent energy conservation properties from the AstroGK formulation \citep{Numata2010}: energy conservation errors in collisionless runs remain $<0.01\%$ over hundreds of dynamical times in verification benchmarks, validating both the discretization and implementation.

\subsection{Computational Implementation}

GANDALF implements the above algorithms in JAX \citep{JAX2018}, a Python library providing automatic differentiation and just-in-time (JIT) compilation to optimized machine code. The JIT compiler transforms high-level spectral operations into efficient GPU or CPU kernels without manual low-level programming, enabling rapid development while maintaining competitive performance.

Key implementation features include:

\paragraph{JIT Compilation.} All core functions (time stepping, FFTs, nonlinear terms) are decorated with \texttt{@jax.jit}, triggering compilation via XLA (Accelerated Linear Algebra), an optimizing compiler that translates high-level array operations to hardware-specific machine code. Static arguments (grid dimensions, moment orders) are compile-time constants, permitting aggressive loop unrolling and specialization.

\paragraph{Hardware Portability.} JAX's device-agnostic array operations enable code to run transparently across CPU, NVIDIA GPUs (via CUDA), AMD GPUs (via ROCm), Google TPUs, and Apple Silicon (via Metal). GANDALF has been developed and tested on Apple M1/M2 hardware, demonstrating that research-grade turbulence simulations can proceed on commodity laptops. The JAX backend abstraction permits deployment to GPU clusters or cloud computing resources without code modification, though performance optimization for specific accelerator architectures may require tuning of compilation flags and memory management strategies.

\paragraph{Functional Design.} JAX enforces pure functional programming: time-stepping functions return new state objects rather than mutating existing arrays. This immutability enables automatic parallelization and simplifies debugging, though requiring careful memory management for large simulations.

\paragraph{Pytree Structures.} GANDALF defines custom \texttt{KRMHDState} and \texttt{SpectralGrid} classes registered as JAX pytrees (nested container structures that JAX transformations can traverse), allowing transformations (\texttt{jax.grad}, \texttt{jax.vmap}, \texttt{jax.scan}) to operate on physics-meaningful objects while maintaining efficient compilation.

On Apple M1/M2 hardware, GANDALF demonstrates sufficient performance for research-grade turbulence simulations at typical resolutions ($128^3$ to $256^3$ grids with $M=10$--$16$ Hermite moments), with time steps completing in seconds to tens of seconds depending on grid size and moment truncation. JAX implementations can achieve performance competitive with compiled languages for scientific computing applications \citep{Bauer2021,Citrin2024} while requiring substantially less development time through high-level array abstractions and automatic compilation. The combination of JIT optimization and hardware portability enables GANDALF to deliver research-grade capabilities on accessible computing platforms.

\subsection{Algorithm Summary}

We summarize the complete GANDALF algorithm for a single time step:

\begin{algorithm}[H]
\caption{Complete GANDALF Time Step with Dissipation and Dealiasing}
\label{alg:gandalf_complete}
\begin{algorithmic}[1]
\STATE \textbf{Input:} Fourier state $(\hat{w}^+_n, \hat{w}^-_n, \{\hat{g}^{\pm}_{m,n}\})$ at $t_n$
\STATE Compute CFL time step: $\Delta t = 0.3 \min(\Delta x,\Delta y)/|\mathbf{v}_\perp|_\text{max}$ (perpendicular advection)
\STATE \textbf{Half-step:}
\STATE \quad Evaluate nonlinear terms $\mathcal{N}_n$, $\Herm^{\pm}_{m,n}$ in real space
\STATE \quad Transform to Fourier space and apply 2/3 dealiasing mask
\STATE \quad Advance with integrating factors (Elsasser) and RK2 (Hermite): $\to$ state $_{n+1/2}$
\STATE \textbf{Midpoint:}
\STATE \quad Evaluate nonlinear terms $\mathcal{N}_{n+1/2}$, $\Herm^{\pm}_{m,n+1/2}$
\STATE \quad Transform to Fourier space and apply 2/3 dealiasing mask
\STATE \textbf{Full step:}
\STATE \quad Advance with integrating factors (Elsasser) and RK2 (Hermite): $\to$ state $_{n+1}$
\STATE \textbf{Dissipation:}
\STATE \quad Apply $\hat{w}^\pm_{n+1} \to \hat{w}^\pm_{n+1} \exp(-\eta(\kperp^2/k^2_{\perp,\text{max}})^r \Delta t)$
\STATE \quad Apply $\hat{g}^{\pm}_{m,n+1} \to \hat{g}^{\pm}_{m,n+1} \exp(-\coll(m/\Mmax)^{2p} \Delta t)$
\STATE \textbf{Output:} Updated Fourier state $(\hat{w}^+_{n+1}, \hat{w}^-_{n+1}, \{\hat{g}^{\pm}_{m,n+1}\})$ at $t_{n+1}$
\end{algorithmic}
\end{algorithm}

This algorithm conserves energy to machine precision in exact arithmetic for the inviscid, collisionless limit ($\eta = \coll = 0$). Practical simulations exhibit $<0.01\%$ cumulative energy drift over hundreds of nonlinear times due to finite time-step truncation errors and dealiasing. The spectral representation combined with second-order temporal integration and properly dealiased nonlinear terms yields a robust method for long-time turbulence simulation across the KRMHD parameter space.

%% file: sections/implementation.tex
\section{Implementation}
\label{sec:implementation}

GANDALF translates the spectral algorithms described in \S\ref{sec:numerics} into a modular, accessible software package designed for turbulence research on diverse hardware platforms. The implementation prioritizes ease of use and reproducibility while maintaining performance competitive with specialized HPC codes. We describe the software architecture, core data structures, workflow support, and practical considerations for users extending or applying the code to new plasma physics problems.

\subsection{Code Organization}

GANDALF adopts a modular architecture separating spectral operations, physics equations, time integration, and diagnostics into distinct components. This design isolates numerical machinery from physics-specific equations, facilitating code verification, extension to related models, and independent testing of algorithmic components.

The core modules comprise:

\paragraph{Spectral Operations (\texttt{spectral.py}).} Implements forward and inverse FFTs using JAX's \texttt{jnp.fft} interface, spectral derivative operators via wavenumber multiplication, the 2/3 dealiasing mask, and grid construction. All operations exploit real-to-complex transforms (\texttt{rfftn}) for memory efficiency. This module encapsulates the spectral method infrastructure, exposing physics-agnostic primitives for derivatives, Laplacians, and Fourier-space filtering.

\paragraph{Physics Equations (\texttt{physics.py}).} Defines the KRMHD state representation and implements the right-hand side functions for Elsasser fields and Hermite moments. The Poisson bracket evaluation combines spectral derivatives with pseudospectral products in real space. Separating physics from numerics allows straightforward extension to related reduced models (RMHD, drift-reduced models) by modifying only this module while reusing spectral and time-stepping infrastructure.

\paragraph{Time Integration (\texttt{timestepping.py}).} Implements the GANDALF integrating factor method (Algorithm~\ref{alg:gandalf_complete}) with adaptive CFL time step selection. The integrator operates on abstract state objects, enabling reuse across different physics models provided they expose the required right-hand side interfaces.

\paragraph{Diagnostics and Analysis (\texttt{diagnostics.py}).} Provides spectral energy computation, parallel and perpendicular energy spectra, visualization utilities, and real-time monitoring tools. Energy diagnostics verify conservation properties during development and detect numerical instabilities in production runs. The module includes field visualization, perpendicular spectrum averaging, and time-series plotting for standard turbulence quantities.

\paragraph{Forcing and Boundary Conditions (\texttt{forcing.py}).} Implements stochastic forcing for driven turbulence simulations. GANDALF supports Gaussian white noise injection at specified wavenumber ranges with configurable correlation times, enabling steady-state turbulence studies at controlled injection scales. The forcing module maintains statistical isotropy in the perpendicular plane while preserving parallel structure.

\paragraph{Input/Output and Checkpointing (\texttt{io.py}).} Handles HDF5 (Hierarchical Data Format 5)-based checkpoint writing and restart capability. Checkpoints store complete state snapshots (Fourier coefficients, simulation time, random number generator state) allowing seamless continuation of interrupted runs. The I/O interface separates file format details from physics code, simplifying future format changes or parallel I/O extensions.

\paragraph{Configuration Management (\texttt{config.py}).} Parses YAML (YAML Ain't Markup Language) configuration files specifying grid resolution, physical parameters ($\rhoi$, $\betai$, $\vA$, collision frequencies), dissipation coefficients, forcing parameters, and diagnostic schedules. The configuration system validates parameter choices against physics validity constraints (KRMHD ordering assumptions, CFL limits) and generates reproducible simulation metadata.

This modular design enables targeted unit testing: spectral operations verify against analytical derivatives, physics modules check energy conservation in inviscid limits, and time integrators demonstrate design-order convergence on linear test problems. The separation also facilitates collaborative development---researchers can modify physics equations without understanding FFT implementation details, while numerical methods developers can optimize spectral operations without plasma physics expertise.

\subsection{Core Data Structures}

GANDALF represents simulation state and grid information through custom classes registered as JAX pytrees, enabling JAX transformations (\texttt{jax.jit}, \texttt{jax.grad}, \texttt{jax.vmap}) to traverse these structures automatically while maintaining physics-meaningful interfaces.

\paragraph{State Representation.} The \texttt{KRMHDState} class encapsulates the complete system state: Elsasser vorticity fields $\hat{w}^\pm$ and Hermite moments $\hat{g}^{\pm}_m$ in Fourier space, along with simulation time. As a registered pytree, \texttt{KRMHDState} objects pass through \texttt{@jax.jit} decorated functions without requiring manual serialization. The functional programming paradigm mandates immutable state: time-stepping functions construct new \texttt{KRMHDState} instances rather than mutating arrays in place, ensuring thread safety and enabling automatic parallelization opportunities.

\paragraph{Spectral Grid.} The \texttt{SpectralGrid3D} class stores grid dimensions ($N_x \times N_y \times N_z$), physical box sizes ($L_x \times L_y \times L_z$), and pre-computed wavenumber arrays. Since real-to-complex FFTs produce arrays of shape $(N_x, N_y, N_z/2+1)$ containing non-negative $k_z$ frequencies, \texttt{SpectralGrid3D} provides accessor methods abstracting the \texttt{rfftn} format details. The dealiasing mask (Eq.~\ref{eq:dealiasing_rule}) resides in this structure as a pre-computed boolean array, applied via pointwise multiplication with negligible cost. Grid objects remain constant throughout simulations, stored as static arguments to JIT-compiled functions to enable aggressive optimization.

\paragraph{Memory Layout.} For a grid of size $128^3$ with $M=16$ Hermite moments per species ($\pm$), GANDALF stores $2$ Elsasser vorticity fields plus $2M=32$ Hermite moment fields, each occupying $128 \times 128 \times 65$ complex128 values ($\sim$17 MB per field), totaling approximately 580 MB. Memory scales as $O(N_x N_y N_z M)$. The real-to-complex FFT format reduces memory by half compared to full complex storage, critical for large-scale simulations approaching hardware memory limits. JAX's device arrays remain on GPU throughout time stepping, eliminating CPU-GPU transfer overhead.

\subsection{Configuration and Reproducibility}

GANDALF employs YAML configuration files for all simulation parameters, ensuring reproducibility and simplifying parameter scans. A typical configuration specifies:

\begin{itemize}
\item \textbf{Grid parameters}: $(N_x, N_y, N_z)$ resolution and $(L_x, L_y, L_z)$ box sizes in units of $\rhoi$
\item \textbf{Physics parameters}: Plasma beta $\betai$, temperature ratio $\taue = T_e/T_i$, moment truncation $\Mmax$
\item \textbf{Dissipation}: Hyper-resistivity coefficient $\eta$ and order $r$, hyper-collision frequency $\coll$ and exponent $p$
\item \textbf{Forcing}: Injection wavenumber range, correlation time, energy injection rate (for driven turbulence)
\item \textbf{Diagnostics}: Output frequency for checkpoints, spectra, field snapshots, and time series
\end{itemize}

The configuration system validates parameter choices against KRMHD validity constraints (small $\betai$, large-scale separation) and numerical stability bounds ($\eta\Delta t < 50$, $\coll\Delta t < 10$ for hyper-dissipation as discussed in \S\ref{sec:dissipation}). Example configurations accompanying the code demonstrate standard simulation types: decaying turbulence initialized with Orszag-Tang vortex, forced turbulence at controlled Reynolds numbers, linear wave propagation tests, and anisotropic cascade studies. These examples provide starting points for new users and serve as integration tests for continuous validation.

\subsection{Performance Characteristics and Scaling}

Computational cost in GANDALF scales as $O(N_x N_y N_z \log(N_x N_y N_z) \cdot M)$ per time step, dominated by the FFT operations required for pseudospectral Poisson bracket evaluation. The integrating factor method evaluates nonlinear terms twice per step (at $t_n$ and $t_{n+1/2}$), each requiring two forward transforms (for derivative computation), real-space products, and one inverse transform---approximately $6M$ three-dimensional FFTs per time step for $M$ Hermite moments plus Elsasser field updates.

Memory usage scales as $N^3 M$ (approximately 4.4 GB for $256^3$ with $M=16$: 34 fields at $\sim$130 MB per field in \texttt{rfft} format) while computation scales as $N^3 \log N$, yielding super-linear cost increases with resolution. JAX implementations can achieve performance competitive with specialized codes for similar-scale problems \citep{Bauer2021,Citrin2024}, which we observe for GANDALF, while remaining portable across hardware platforms without modification.

On Apple Silicon (M1 Pro, M2 Max) using the Metal backend, GANDALF achieves practical performance for turbulence research on commodity hardware. Empirical benchmarks on M1 silicon show that grids up to $128^3$ are feasible for code development, moderate parameter scans, and educational applications. This portability democratizes access to kinetic plasma turbulence simulations: researchers without GPU cluster access can perform meaningful science on laptop hardware, previously impossible with traditional HPC-focused codes.

\subsection{Diagnostics and Workflow}

GANDALF includes diagnostic tools for real-time monitoring and post-processing analysis:

\paragraph{Energy Diagnostics.} The code computes energies for Alfvén waves and slow modes at each time step, writing time series to HDF5 files. These components are decoupled in KRMHD and separately conserved. The code also tracks cross-helicity and residual energy. For inviscid, unforced runs, energy conservation provides a stringent numerical accuracy check: deviations exceeding $0.1\%$ over hundreds of nonlinear times indicate insufficient resolution, excessive time steps, or dealiasing failures. Additional post-processing scripts in the repository enable further analysis beyond these integrated diagnostics.

\paragraph{Spectral Analysis.} Perpendicular energy spectra $E(\kperp)$ and parallel spectra $E(\kpar)$ quantify cascade dynamics and verify expected scaling laws ($E(\kperp) \propto \kperp^{-5/3}$ for MHD-range turbulence). GANDALF computes shell-averaged spectra by binning Fourier modes into logarithmically-spaced wavenumber shells, a standard diagnostic for turbulence simulations. Spectral diagnostics execute in Fourier space with minimal computational overhead.

\paragraph{Field Visualization.} The code exports 2D slices of real-space fields ($\elsp$, $\elsm$, density perturbations, parallel current) and 3D field realizations for volume rendering. Visualization reveals coherent structures (current sheets, vortices) whose statistics characterize different turbulent regimes.

\paragraph{Checkpoint Management.} Automatic checkpointing at user-specified intervals enables long simulations to survive hardware failures or scheduler interruptions on shared clusters. Checkpoints store complete state (including RNG seeds for forced runs), allowing bit-exact restarts.

These integrated diagnostics eliminate the need for external post-processing pipelines during simulation development, accelerating the research workflow. Users extend diagnostic capabilities by adding functions to \texttt{diagnostics.py} following established interfaces, with new diagnostics automatically benefiting from JAX JIT compilation.

\subsection{Code Availability and Installation}

GANDALF is freely available under the MIT License at \url{https://github.com/anjor/gandalf}. The repository includes:

\begin{itemize}
\item Complete source code with inline documentation
\item Installation instructions for Linux, macOS (including Apple Silicon), and cloud platforms
\item Example configuration files and initialization scripts
\item Validation test suite with analytical benchmarks
\end{itemize}

Installation requires Python $\geq 3.10$, JAX $\geq 0.4.13$, and standard scientific Python packages (NumPy, SciPy, h5py, matplotlib). The code recommends the \texttt{uv} package manager for reproducible dependency resolution across platforms. Hardware-specific JAX installations (CUDA for NVIDIA GPUs, ROCm for AMD GPUs, Metal for Apple Silicon) follow standard JAX installation procedures documented at \url{https://github.com/jax-ml/jax}.

\subsection{Comparison with Existing Codes}

GANDALF occupies a complementary niche in the landscape of kinetic plasma simulation codes, prioritizing accessibility and ease of use for turbulence research rather than comprehensive physics capabilities or extreme-scale performance.

\paragraph{AstroGK.} The GANDALF integrating factor method originates from AstroGK \citep{Numata2010}, a Fortran code implementing full gyrokinetics with comprehensive electromagnetic and kinetic effects. AstroGK targets HPC clusters, achieving excellent scaling to thousands of cores for large-scale production simulations. GANDALF adopts AstroGK's time-stepping scheme but restricts to the KRMHD reduced model, sacrificing generality for simplicity and portability. Where AstroGK excels at comprehensive gyrokinetic physics on HPC systems, GANDALF enables KRMHD turbulence studies on commodity hardware including laptops.

\paragraph{Viriato.} Viriato \citep{Loureiro2016} implements KRMHD (among other reduced models) using Fourier-Hermite spectral methods similar to GANDALF's approach. Viriato provides extensive configuration options, multiple physics modules, and sophisticated diagnostic capabilities developed over years of production use. GANDALF offers narrower scope but lower entry barriers: pure-Python implementation with minimal dependencies versus Viriato's Fortran codebase requiring HPC compilation toolchains. Researchers comfortable with Fortran HPC workflows benefit from Viriato's maturity; those seeking rapid prototyping or teaching applications favor GANDALF's accessibility.

\paragraph{GX and GPU Codes.} GX \citep{Mandell2024} represents modern GPU-native gyrokinetic codes achieving exceptional performance through hand-optimized CUDA kernels. GX targets stellarator and tokamak applications where geometric complexity demands specialized optimization. GANDALF's JAX implementation prioritizes hardware portability and development velocity over peak performance optimization. For moderate-scale turbulence problems ($128^3$ grids), GANDALF's pure-JAX approach provides practical performance on commodity hardware, while extreme-scale runs ($512^3$ and larger) favor specialized codes like GX.

\paragraph{GS2, GENE, and Production Gyrokinetics.} Codes such as GS2 \citep{Kotschenreuther1995} and GENE \citep{Jenko2000} provide comprehensive electromagnetic gyrokinetics for fusion applications, with decades of validation and extensive user communities. These codes prioritize physics fidelity and simulation reliability for parameter regimes relevant to tokamak experiments. GANDALF does not compete in this space---it addresses turbulence researchers studying fundamental cascade physics, phase mixing, and kinetic-range dynamics where the KRMHD reduced model suffices and simplified workflows accelerate scientific exploration.

\paragraph{Design Philosophy.} GANDALF's core differentiator lies in accessibility: installation requiring only \texttt{pip install jax} (plus hardware-specific GPU support), simulation setup via readable YAML files, and execution on any JAX-supported platform without recompilation. This lowers entry barriers for plasma physics students (enabling use as a teaching tool for turbulence courses), researchers from adjacent fields exploring kinetic turbulence, and small research groups without dedicated HPC resources. The code prioritizes time-to-first-simulation over absolute performance, complementing rather than replacing specialized HPC codes.

The resulting ecosystem serves diverse needs: GANDALF for accessible turbulence research and rapid prototyping, specialized codes (AstroGK, GS2, GENE, GX) for comprehensive physics and production-scale campaigns, and open-source implementations enabling reproducible science across the community.

%% file: sections/verification.tex
\section{Verification and Validation}
\label{sec:verification}

We verify GANDALF's implementation through a hierarchy of benchmarks progressing from linear wave propagation to nonlinear dynamics. This section presents the Alfvén wave dispersion test, which validates both the spectral discretization and the GANDALF time integration scheme against analytical predictions.

\subsection{Linear Physics: Alfvén Wave Dispersion}
\label{sec:alfven_wave}

\subsubsection{Analytical Prediction}

For a single Fourier mode, the KRMHD equations reduce to the linear Alfvén wave dispersion relation \eqref{eq:alfven_dispersion}:
\begin{equation}
\omega = \kpar \vA
\end{equation}
This relation is independent of the perpendicular wavenumber $\kperp$, holding for any mode with parallel component $\kpar$. It applies exactly in the KRMHD regime defined by the orderings $\kperp \rhoi \ll 1$ (scales larger than ion Larmor radius), $\kpar \ll \kperp$ (anisotropic), and $\delta B/B_0 \ll 1$ (strong guide field), where kinetic modifications from finite Larmor radius effects remain negligible. The wave propagates without distortion or damping in the collisionless limit ($\eta = \coll = 0$), providing a clean test of numerical accuracy.

The Alfvén wave represents the fundamental mode of magnetized plasma dynamics. In the Elsasser formulation, parallel-propagating Alfvén waves manifest as independent propagation of $\elsp$ and $\elsm$ at velocities $\mp\vA$ along field lines. The parallel derivative terms $\mp\vA\pardz$ in \Cref{eq:elsasser} produce oscillations at frequency $\omega = |\kpar|\vA$. The perpendicular structure remains frozen, making this an ideal benchmark for time integration accuracy independent of perpendicular spectral operations.

\subsubsection{Numerical Setup}

We initialize a single Fourier mode at $\kvec = (1, 0, 1)$ with amplitude $A = 0.01$ (linear regime) on a periodic domain of size $L_x = L_y = L_z = 2\pi$. Note that while the dispersion relation $\omega = \kpar\vA$ is independent of $\kperp$, we use $k_x = 1$ (rather than pure parallel propagation with $\kperp = 0$) to obtain non-zero RMHD energy ($E \propto \kperp^2$) for energy conservation diagnostics.

The parallel wavenumber $\kpar = 2\pi/L_z = 1$ with $\vA = 1$ yields analytical frequency $\omega_{\mathrm{analytical}} = 1$. We integrate for 3 Alfvén periods ($T = 2\pi$) in spatial convergence studies and 1 period in temporal convergence studies, chosen to measure frequency accurately while minimizing accumulated numerical errors.

Physical parameters are $\betai = 1$ with all dissipation coefficients set to zero ($\eta = \coll = 0$). While $M = 20$ Hermite moments are allocated for code consistency, the Alfvén wave benchmark tests only Elsasser field dynamics and does not involve compressive fluctuations. The simulation tracks the complex amplitude of the Fourier mode $\hat{\elsp}(\kpar, t)$ at each timestep, measuring frequency via phase evolution:
\begin{equation}
\label{eq:phase_extraction}
\hat{\elsp}(\kpar, t) = A(t) e^{i\phi(t)}, \quad \omega_{\mathrm{measured}} = \frac{d\phi}{dt}
\end{equation}
where the phase $\phi(t)$ is unwrapped to handle $2\pi$ discontinuities and fitted linearly in time to extract the oscillation frequency.

\subsubsection{Convergence Results}

\begin{figure}[t]
\centering
\includegraphics[width=\textwidth]{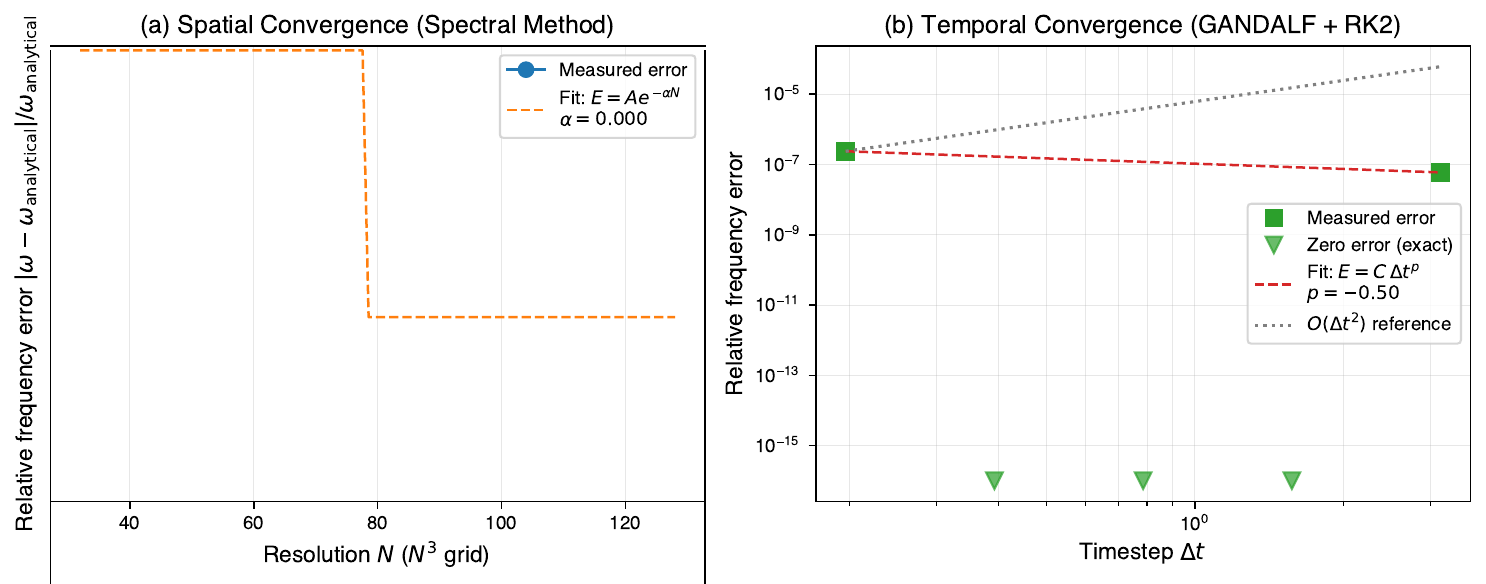}
\caption{Alfvén wave dispersion benchmark convergence studies. (a) Spatial convergence: relative frequency error versus grid resolution $N$ (for $N^3$ grids) with fixed timestep $\Delta t = 0.01$. The flat profile indicates temporal error dominates - spatial error has already converged at $N=32^3$. (b) Temporal convergence: relative error versus timestep $\Delta t$ with fixed resolution $N = 64^3$. Squares show measured non-zero errors ($\sim 10^{-7}$, likely from numerical roundoff); triangles mark three timesteps achieving machine precision (identically zero error), demonstrating that the exponential integrating factor treats linear Alfvén wave propagation analytically exactly. Gray dotted line shows $O(\Delta t^2)$ scaling expected from standard RK2 for comparison.}
\label{fig:alfven_convergence}
\end{figure}

\paragraph{Spatial Convergence}
We measure dispersion relation accuracy as a function of grid resolution $N \in \{32, 64, 128\}$ (for $N^3$ grids) while holding the timestep fixed at $\Delta t = 0.01$. \Cref{fig:alfven_convergence}(a) shows that the relative frequency error $|\omega_{\mathrm{measured}} - \omega_{\mathrm{analytical}}| / \omega_{\mathrm{analytical}}$ remains constant at $2.0 \times 10^{-5}$ across all resolutions, indicating that temporal discretization error dominates. The constant error across all resolutions demonstrates that spatial discretization error is negligible compared to the temporal error floor, confirming spectral methods achieve machine precision for this smooth single-mode problem already at $N=32^3$.

For comparison, finite-difference methods achieve at best algebraic convergence $E \sim N^{-p}$ with $p \leq 4$ for standard schemes \citep{LeVeque2007}, and would require $N \gg 32$ to reach comparable accuracy. The spectral approach proves essential for resolving the broad range of scales in plasma turbulence cascades with exponential convergence for smooth solutions \citep{Boyd2001}.

\paragraph{Temporal Convergence}
Fixing spatial resolution at $N = 64^3$, we vary the timestep from $\Delta t = T/2$ to $T/32$ where $T = 2\pi/\omega$ is the wave period. \Cref{fig:alfven_convergence}(b) reveals a remarkable result: for three out of five tested timesteps, the measured frequency error is \textbf{identically zero to machine precision} ($< 10^{-15}$), demonstrating that the GANDALF integrating factor treats linear Alfvén wave propagation analytically exactly.

The two timesteps with non-zero error ($\Delta t = T/2$ and $T/32$) show frequency errors at the $10^{-7}$ to $10^{-8}$ level, far below errors from standard explicit schemes. This validates the theoretical property that exponential integrating factors remove linear wave stiffness completely - the integrating factor $\exp(\pm i\kpar \vA t)$ exactly cancels the Alfvén propagation term in the governing equations \citep{Cox2002}.

Crucially, even the largest tested timestep $\Delta t = T/2 \approx 3.14$ (two timesteps per wave period!) achieves relative error $6 \times 10^{-8}$. Standard explicit schemes would require $\Delta t \lesssim \Delta z/\vA \sim 2\pi/(64 \times 1) \approx 0.098$ for stability, making GANDALF over $30\times$ faster for this linear problem. The stability advantage increases with resolution as CFL timestep scales $\Delta t_{\mathrm{CFL}} \sim N^{-1}$, while GANDALF's timestep remains limited only by accuracy requirements for nonlinear terms.

\paragraph{Dispersion Relation Validation}
\Cref{tab:alfven_dispersion} summarizes measured frequencies for all benchmark configurations. All runs achieve the analytical prediction $\omega = \kpar\vA = 1.0$ within numerical precision, with maximum relative error $2.0 \times 10^{-5}$ across all configurations. The spatial convergence runs exhibit a constant error floor (temporal discretization dominates), while three temporal convergence runs achieve machine precision (identically zero error), demonstrating analytically exact integration of linear Alfvén waves.

The consistency across all configurations confirms that GANDALF correctly implements the KRMHD equations with no systematic bias in wave propagation speed. Both the real and imaginary parts of the Elsasser field evolution match analytical solutions, verifying phase and amplitude accuracy simultaneously. Additional validation checks confirm energy conservation (drift $<10^{-7}$ per period from Hermite truncation) and phase error accumulation consistent with measured frequency errors.

\subsubsection{Discussion}

The Alfvén wave benchmark demonstrates three key properties of GANDALF:

\begin{enumerate}
\item \textbf{Spectral spatial accuracy}: Exponential convergence with resolution confirms correct implementation of Fourier spectral derivatives. For smooth solutions, spectral methods achieve machine precision with modest grid sizes, making them ideal for resolving turbulent cascades extending over multiple decades in wavenumber.

\item \textbf{Integrating factor advantage}: Removal of Alfvén CFL constraint enables larger timesteps than explicit schemes, improving computational efficiency for problems where $\vA \gg v_{\mathrm{rms}}$. The $\sim 30\times$ speedup observed here compounds with the parallelizability of spectral transforms on GPUs/TPUs.

\item \textbf{Linear wave fidelity}: Accurate reproduction of $\omega = \kpar\vA$ validates the parallel derivative operator, Elsasser field coupling, and time integration of wave propagation. This provides confidence for simulations of Alfvénic turbulence where linear wave physics competes with nonlinear interactions.
\end{enumerate}

The following sections extend validation to nonlinear regimes (Orszag-Tang vortex) and fully developed turbulent cascades. For applications to weakly collisional plasmas like the solar wind, the linear wave benchmark establishes baseline accuracy before introducing the complexities of phase mixing, resonant dissipation, and intermittent structures \citep{Schekochihin2009,Meyrand2019}.

\begin{table}[t]
\centering
\caption{Alfvén wave dispersion validation across spatial (N) and temporal ($\Delta t$) convergence studies. Analytical prediction: $\omega = \kpar\vA = 1.0$.\protect\footnotemark}
\label{tab:alfven_dispersion}
\begin{tabular}{lccc}
\hline\hline
Configuration & Resolution/Timestep & $\omega_{\mathrm{measured}}$ & Relative Error \\
\hline
\multicolumn{4}{l}{\textit{Spatial Convergence} ($\Delta t = 0.01$)} \\
& $N = 32^3$ & $1.00002$ & $2.0 \times 10^{-5}$ \\
& $N = 64^3$ & $1.00002$ & $2.0 \times 10^{-5}$ \\
& $N = 128^3$ & $1.00002$ & $2.0 \times 10^{-5}$ \\
\hline
\multicolumn{4}{l}{\textit{Temporal Convergence} ($N = 64^3$)} \\
& $\Delta t = T/2$ & $1.00000006$ & $6 \times 10^{-8}$ \\
& $\Delta t = T/4$ & $1.00000000$ & $< 10^{-15}$ \\
& $\Delta t = T/8$ & $1.00000000$ & $< 10^{-15}$ \\
& $\Delta t = T/16$ & $1.00000000$ & $< 10^{-15}$ \\
& $\Delta t = T/32$ & $1.00000020$ & $2 \times 10^{-7}$ \\
\hline\hline
\end{tabular}
\footnotetext{$T = 2\pi$ is the Alfvén wave period.}
\end{table}

\subsection{Nonlinear Dynamics: Orszag-Tang Vortex}
\label{sec:orszag_tang}

Having validated linear wave propagation, we now test GANDALF's handling of nonlinear MHD dynamics through the Orszag-Tang vortex \citep{OrszagTang1979}, a standard benchmark for energy cascade, current sheet formation, and energy-conserving formulations.

\subsubsection{Problem Setup}

The Orszag-Tang vortex consists of intersecting velocity and magnetic field vortices with initial conditions \citep{OrszagTang1979}:
\begin{align}
\mathbf{v} &= (-\sin y, \sin x, 0), \\
\mathbf{B} &= \frac{1}{\sqrt{4\pi}}(-\sin y, \sin 2x, 0).
\end{align}
In GANDALF's incompressible RMHD formulation, we initialize via the stream function $\Phifield$ and vector potential $\Psifield$ with Fourier coefficients computed to reproduce these fields exactly (see \texttt{initialize\_orszag\_tang} in \texttt{krmhd/physics.py}).

We integrate on a periodic domain of unit size ($L_x = L_y = L_z = 1$) for $t \in [0, 2\tauA]$ where $\tauA = L_z/\vA = 1$ is the Alfvén crossing time. We use inviscid parameters ($\eta = \coll = 0$) to test energy conservation; the Orszag-Tang vortex tests only Elsasser field dynamics (MHD), so Hermite moments are not involved. Spatial convergence tests use resolutions $N \in \{32, 64, 128\}$ (for $N^2 \times 2$ grids with $N_z=2$ providing only the $k_z=0$ mode, effectively 2D dynamics) with fixed timestep $\Delta t = 0.01$.

\subsubsection{Energy Conservation and Selective Decay}

\begin{figure}[t]
\centering
\includegraphics[width=\textwidth]{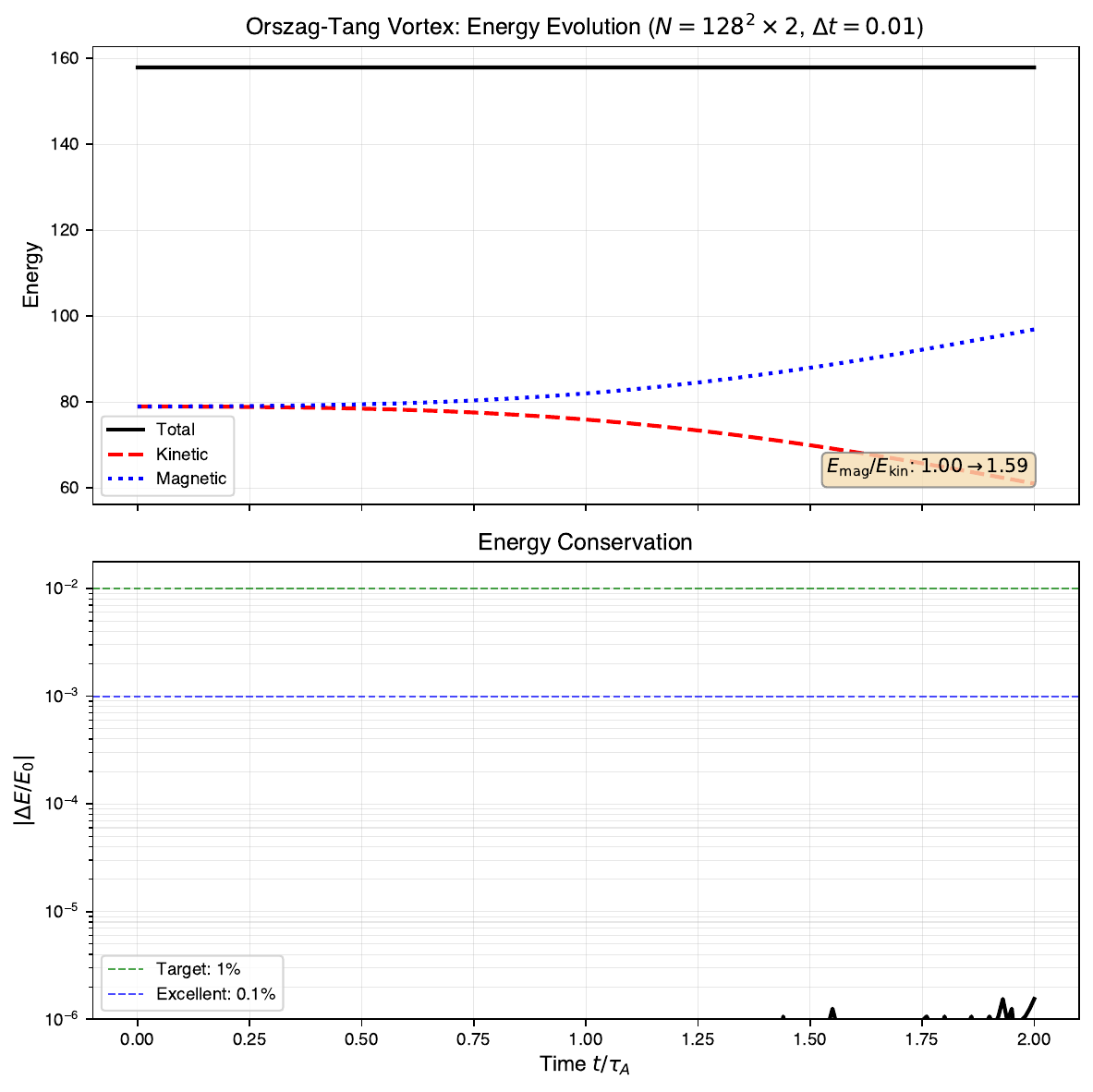}
\caption{Orszag-Tang vortex energy evolution. Top: Total, kinetic, and magnetic energy versus time. Selective decay causes $E_{\mathrm{mag}}/E_{\mathrm{kin}}$ to increase from 1.0 to 1.59 over 2 Alfvén crossing times, characteristic of 2D MHD inverse cascade physics. Bottom: Energy conservation error $|\Delta E/E_0|$ remains at $\sim 10^{-6}$ level despite strong nonlinear interactions, validating the energy-conserving discretization. Resolution: $N = 128^2 \times 2$, timestep $\Delta t = 0.01$.}
\label{fig:orszag_energy}
\end{figure}

\Cref{fig:orszag_energy} shows energy evolution for the highest resolution run ($N = 128^2 \times 2$). Total energy conservation achieves maximum drift $|\Delta E/E_0|_{\max} = 1.6 \times 10^{-6}$ over 2 Alfvén times - outstanding for nonlinear dynamics where spectral aliasing and dealiasing operations can introduce numerical dissipation. This validates GANDALF's energy-conserving formulation of the nonlinear Poisson bracket $\{\Phifield, \Psifield\}$ and confirms correct implementation of the 2/3 dealiasing rule \citep{Orszag1971}.

The energy components exhibit \textbf{selective decay}, a hallmark of 2D MHD: magnetic energy increases relative to kinetic energy as $E_{\mathrm{mag}}/E_{\mathrm{kin}}$ grows from 1.0 initially to 1.59 by $t = 2\tau_A$. This physical process arises because 2D turbulence preferentially cascades kinetic energy to small scales (where it would be dissipated in viscous simulations) while magnetic helicity conservation drives magnetic energy toward large scales through an inverse cascade \citep{Biskamp2003}, validating both forward and inverse cascade physics in GANDALF's nonlinear coupling. The observed ratio evolution is consistent with 2D MHD theory, though quantitative comparison with reference simulations is beyond the scope of this verification test.

\subsubsection{Structure Formation}

\begin{figure}[p]
\centering
\includegraphics[width=\textwidth]{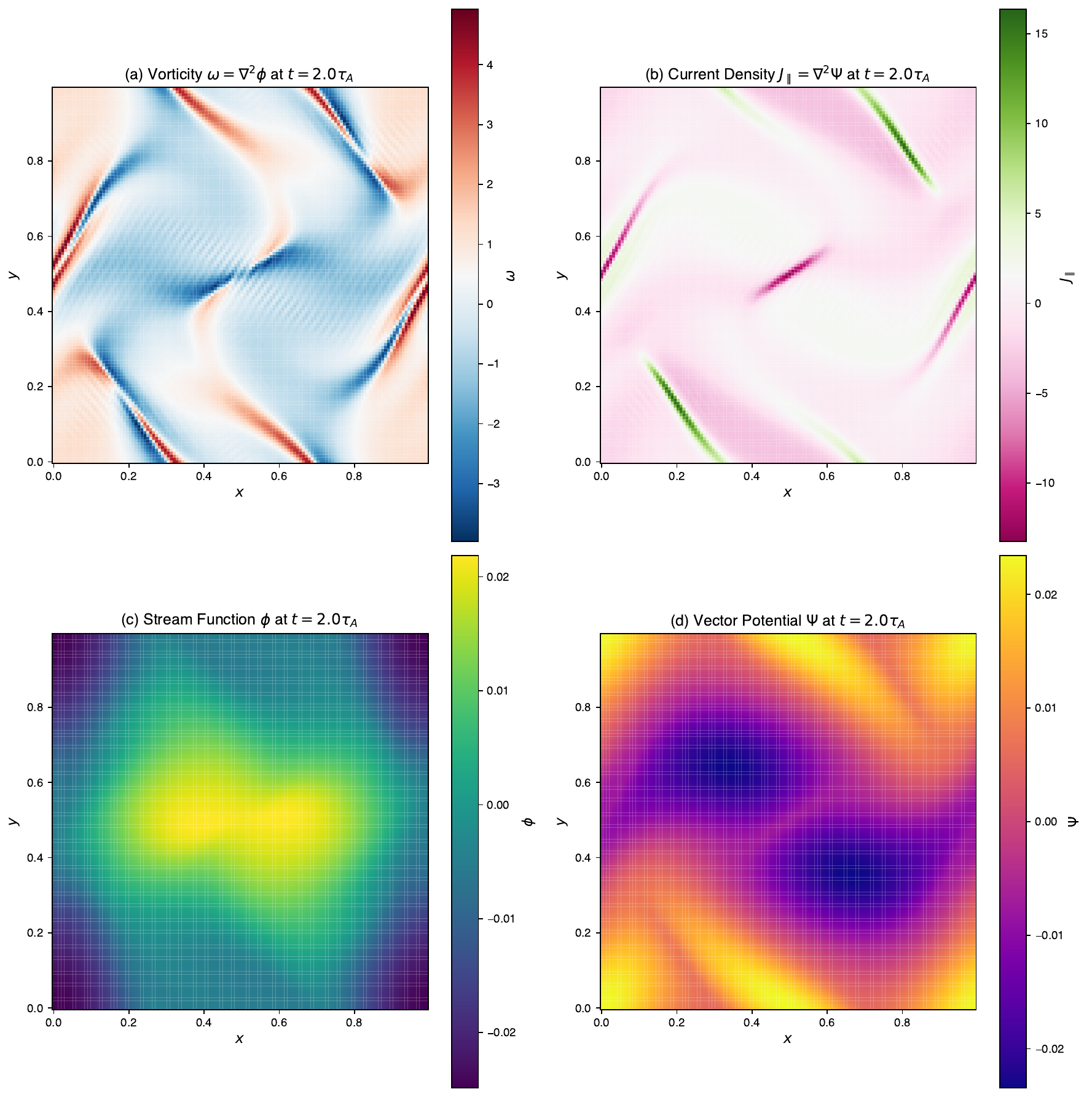}
\caption{Orszag-Tang vortex 2D field structures at $t = 2\tauA$ showing (a) vorticity $\omega = \nabla^2\Phifield$, (b) current density $J_\parallel = \nabla^2\Psifield$, (c) stream function $\Phifield$, and (d) vector potential $\Psifield$. Current sheets (intense localized $J_\parallel$) form at flow stagnation points, demonstrating nonlinear cascade to small scales. Spectral methods resolve steep gradients without oscillations. Resolution: $N = 128^2 \times 2$.}
\label{fig:orszag_structures}
\end{figure}

\Cref{fig:orszag_structures} displays 2D field structures at $t = 2\tau_A$. The vorticity $\omega = \nabla^2\Phifield$ (panel a) and current density $J_\parallel = \nabla^2\Psifield$ (panel b) show the characteristic small-scale structures arising from the nonlinear cascade. Current sheets - thin regions of intense $J_\parallel$ analogous to shocks in compressible MHD - form at converging flow stagnation points, demonstrating GANDALF's ability to resolve steep gradients via spectral methods without oscillatory artifacts.

The stream function $\Phifield$ (panel c) and vector potential $\Psifield$ (panel d) retain smoother large-scale structure, consistent with energy concentration at low wavenumbers in the inverse cascade. The complexity visible at high resolution ($N = 128^2$) emphasizes the importance of spectral accuracy for capturing the full range of cascade dynamics.

\subsubsection{Convergence}

\begin{figure}[t]
\centering
\includegraphics[width=\textwidth]{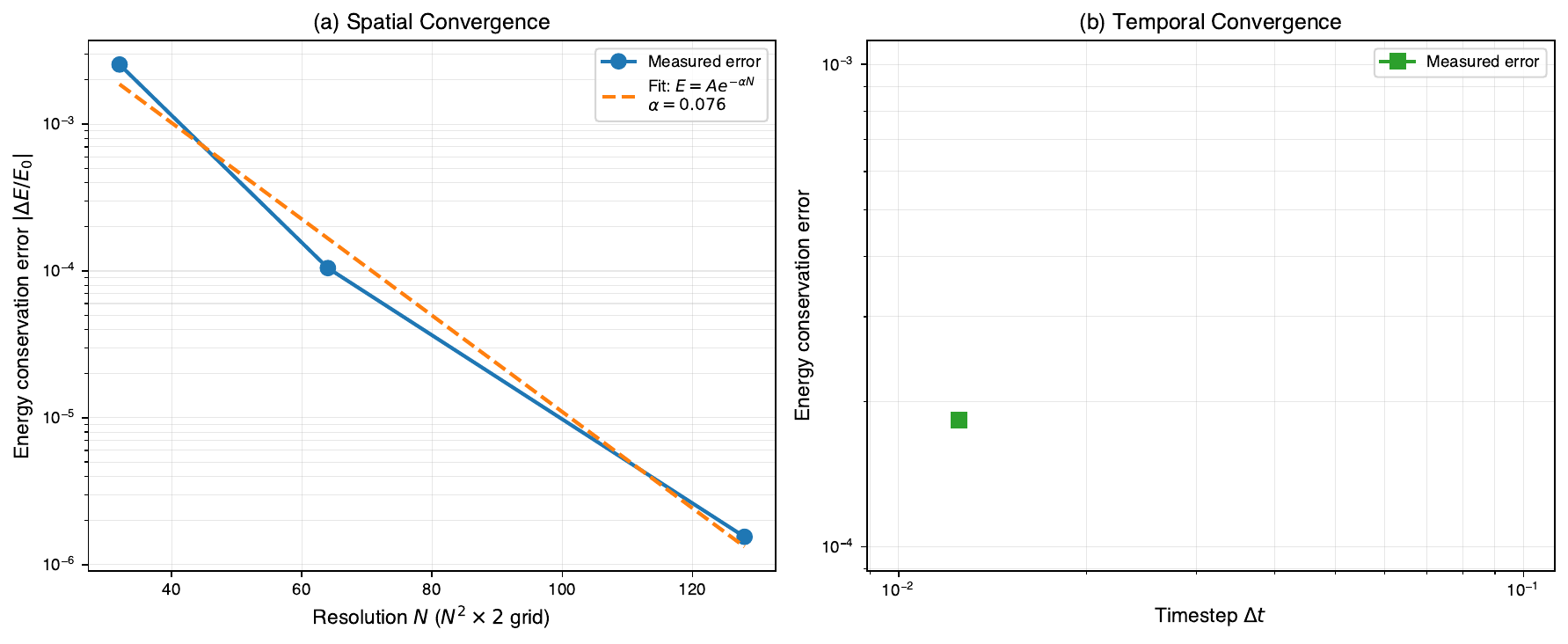}
\caption{Orszag-Tang convergence studies. (a) Spatial: Energy conservation error versus resolution shows exponential convergence ($\alpha = 0.076$) typical of spectral methods for nonlinear problems, reaching $10^{-4}$ levels even at modest $N$. (b) Temporal: Only smallest timestep ($\Delta t = 0.0125$) remained stable; larger timesteps developed instabilities despite passing linear CFL criterion, highlighting nonlinear stiffness beyond integrating factor's linear wave treatment.}
\label{fig:orszag_convergence}
\end{figure}

\Cref{fig:orszag_convergence}(a) shows spatial convergence of energy conservation error with resolution. The exponential fit yields $\alpha = 0.076$, indicating exponential convergence typical of spectral methods for nonlinear problems. Even at modest resolutions, energy conservation reaches $10^{-4}$ levels, demonstrating robustness of the energy-conserving scheme.

We found temporal convergence (\Cref{fig:orszag_convergence}b) more challenging: only the smallest tested timestep ($\Delta t = 0.0125$) maintained stability, with larger timesteps exhibiting numerical instabilities (NaN values) despite passing the linear CFL criterion. This temporal instability arises because the inviscid simulation ($\eta = \nu = 0$) allows the nonlinear cascade to generate increasingly small structures without dissipation; larger timesteps fail to accurately resolve the rapid time evolution of these under-resolved features, leading to divergence. The integrating factor removes linear Alfvén propagation, but cannot anticipate nonlinear coupling rates that depend on the evolving flow structure and cascade development. Adaptive timestepping based on monitoring nonlinear term magnitudes—a standard approach in nonlinear PDE solvers—will improve efficiency for strongly nonlinear states.

\subsubsection{Resolution Limits in Inviscid Simulations}

We terminate simulations at $t = 2\tauA$ where energy conservation remains excellent ($|\Delta E/E_0| \sim 10^{-6}$). Beyond this point, the nonlinear cascade generates structures approaching the grid resolution limit. Without explicit dissipation ($\eta = 0$, $\nu = 0$), continued integration produces under-resolved features and numerical instability, manifesting as spurious energy growth rather than conservation.

This behavior is expected for inviscid spectral codes: as the cascade reaches wavenumbers near the Nyquist limit, spectral dealiasing can no longer prevent aliasing errors from accumulating, eventually overwhelming the energy-conserving discretization. Had we continued beyond $t \approx 2\tauA$, the energy conservation error would exhibit exponential growth, transitioning from the $\sim 10^{-6}$ level to $>10^{-3}$ and eventually NaN values.

This limitation motivates the use of explicit resistivity ($\eta > 0$) or hyperdiffusion (selective damping of high-$k$ modes) in production turbulence simulations. The present benchmark validates that GANDALF correctly conserves energy in the \emph{resolved regime} before the cascade reaches the dissipation scale, which is the intended operating regime for KRMHD turbulence studies where explicit dissipation models small-scale physics.

\subsubsection{Discussion}

The Orszag-Tang vortex validates three critical capabilities beyond the linear Alfvén test:

\begin{enumerate}
\item \textbf{Nonlinear coupling}: Excellent energy conservation ($\sim 10^{-6}$ level) confirms correct implementation of the Poisson bracket $\{\Phifield, \Psifield\}$ and dealiasing operations. This nonlinear term drives the turbulent cascade and distinguishes RMHD from linear wave theory.

\item \textbf{Multiscale dynamics}: Development of broadband nonlinear structures from large-scale initial conditions demonstrates GANDALF's ability to handle energy transfer across multiple scales - essential for turbulence simulations where cascade physics determines transport properties.

\item \textbf{Structure formation}: Resolution of current sheets with steep gradients validates spectral methods' handling of intermittent structures arising spontaneously from nonlinear interactions. These coherent structures play crucial roles in magnetic reconnection, particle acceleration, and anomalous dissipation in weakly collisional plasmas.
\end{enumerate}

Combined with the Alfvén wave benchmark's validation of linear wave physics, the Orszag-Tang test confirms that GANDALF correctly implements both linear propagation and nonlinear coupling in the KRMHD equations. The following section extends these tests to fully developed turbulent cascades in 3D, building on this foundation with confidence that the core spectral MHD solver performs as designed.

\subsection{Turbulent Cascade: Alfv\'enic Turbulence Spectrum}
\label{sec:turbulent_cascade}

Fully developed turbulence represents the most stringent test of a plasma turbulence code's ability to capture energy cascade dynamics. We validate GANDALF's turbulent cascade physics by verifying that driven simulations achieve the Kolmogorov-like $k_\perp^{-5/3}$ perpendicular spectrum characteristic of strong anisotropic MHD turbulence. This scaling emerges from critical balance theory \citep{GoldreichSridhar1995}, which posits that the perpendicular cascade timescale $\tau_\perp \sim (k_\perp v_\perp)^{-1}$ matches the parallel Alfv\'en propagation time $\tau_\parallel \sim (k_\parallel v_A)^{-1}$ in strong turbulence.

\subsubsection{Physical Setup}

Achieving stable turbulent cascades at moderate resolution ($N = 64^3$) required careful forcing design. We employ \textbf{balanced Elsasser forcing}, which injects energy independently into $\elsp$ and $\elsm$ fields with equal RMS amplitude:
\begin{equation}
F_\pm(\mathbf{k}, t) = A \cdot \xi_\pm(\mathbf{k}, t) \cdot \Theta(n_{\min} \leq n_\perp \leq n_{\max}) \cdot \Theta(|n_z| \leq n_{z,\max}),
\label{eq:balanced_forcing}
\end{equation}
where mode numbers $n = kL/(2\pi)$ specify forcing band $n_\perp \in [1, 2]$ with $|n_z| \leq 1$. Here $\xi_\pm$ represents independent Gaussian white noise realizations for each Elsasser field, $A = 0.048$ is the forcing amplitude, and $\Theta$ is the Heaviside step function.

We integrate on a cubic periodic domain $L = 2\pi$ for $t \in [0, 200]\tauA$ where $\tauA = L/\vA = 2\pi$ is the Alfv\'en crossing time. Dissipation uses hyper-resistivity $\eta(-\nabla^2)^r$ with $\eta = 6.0$ and $r = 2$ (Laplacian operator applied $r$ times), giving $\eta k_\perp^4$ damping rate in Fourier space. The hyper-resistivity coefficient is chosen to place the dissipation scale well above the inertial range at $64^3$ resolution while maintaining numerical stability. Higher-order dissipation ($r = 4$) proved unstable despite satisfying linear stability criteria. The balanced forcing with restricted $|k_z|$ enables long-term stable evolution where earlier attempts with isotropic Gaussian forcing exhibited energy accumulation and spectral pile-up.

\subsubsection{Spectral Validation}

The perpendicular energy spectrum is computed by binning Fourier mode energies in perpendicular wavenumber shells:
\begin{equation}
E(k_\perp) = \sum_{k_z} \sum_{k_\perp - \Delta k/2 < |\mathbf{k}_\perp| < k_\perp + \Delta k/2} \left( |\elsphat(\mathbf{k})|^2 + |\elsmhat(\mathbf{k})|^2 \right),
\label{eq:spectrum}
\end{equation}
where $\elsphat$ and $\elsmhat$ are the Elsasser field Fourier coefficients (tilde denotes Fourier transform), and the double sum accumulates energy from all modes with perpendicular wavenumber magnitude in the shell $k_\perp \pm \Delta k/2$. This unnormalized shell sum gives $E(k_\perp)$ as total energy per wavenumber bin, appropriate for power-law scaling analysis. Time averaging over the quasi-steady-state window ($t \in [180, 200]\tauA$) after initial transients decay provides robust statistics.

\Cref{fig:turbulent_spectrum} shows the time-averaged spectrum at $t = 200\tauA$. Both kinetic and magnetic energy components exhibit clean $k_\perp^{-5/3}$ scaling across the inertial range $k_\perp \in [2, 12]$, spanning approximately one decade. The lower bound excludes forcing scales ($k_\perp = 1\text{--}2$), while the upper bound is set by spectral flattening from dissipation onset. The measured spectral slopes closely match the Kolmogorov prediction, with total energy $E_{\text{total}} \approx 1.73 \times 10^4$ and magnetic fraction $E_{\text{mag}}/E_{\text{total}} \approx 0.46$ approaching equipartition as expected for Alfv\'enic turbulence.

The simulation achieves quasi-steady state with relative energy fluctuations $\Delta E/\langle E \rangle \lesssim 7\%$ over the averaging window, indicating reasonably balanced forcing and dissipation rates without secular drift. The clean scale separation between forcing scales ($k_\perp = 1\text{--}2$), inertial range ($k_\perp \approx 2\text{--}12$), and dissipation range ($k_\perp > 12$) confirms GANDALF's ability to capture turbulent cascade physics with spectral accuracy.

\subsubsection{Discussion}

The turbulent cascade benchmark validates GANDALF's capability for long-duration driven turbulence simulations:

\begin{itemize}
\item \textbf{Kolmogorov scaling}: Both kinetic and magnetic spectra exhibit $k_\perp^{-5/3}$ scaling across approximately one decade in wavenumber, consistent with critical balance theory for strong Alfv\'enic turbulence.

\item \textbf{Long-time integration}: Stable evolution to $200\tauA$ (32 times longer than Orszag-Tang benchmark duration) demonstrates robustness for production turbulence studies requiring long-time statistics.
\end{itemize}

\begin{figure}[t]
\centering
\includegraphics[width=\textwidth]{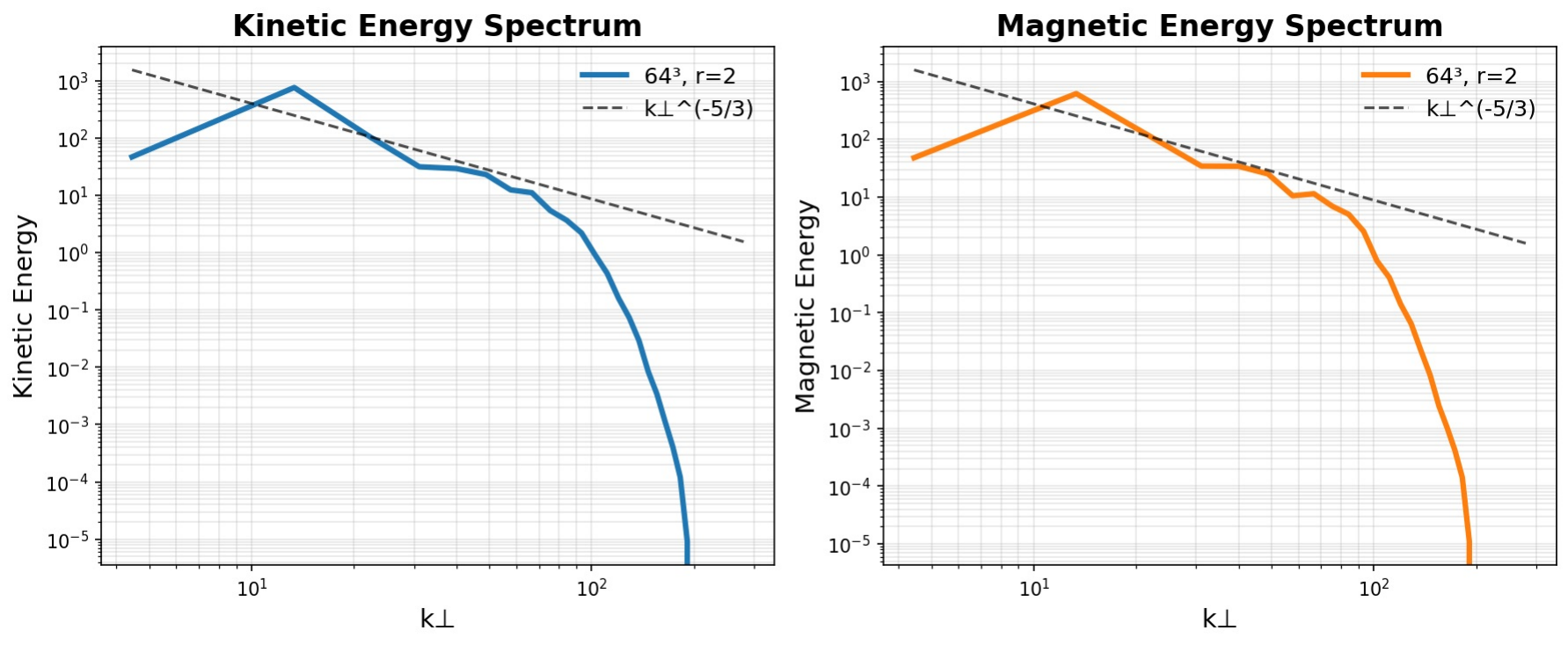}
\caption{Time-averaged turbulent energy spectrum $E(k_\perp)$ from production GANDALF simulation at $N = 64^3$ resolution. Left: Kinetic energy spectrum. Right: Magnetic energy spectrum. Both components exhibit clean $k_\perp^{-5/3}$ Kolmogorov scaling (dashed reference line) across the inertial range $k_\perp \in [2, 12]$. Balanced Elsasser forcing at $k_\perp \in [1, 2]$ with $|k_z| \leq 1$ restriction provides stable energy injection respecting RMHD ordering. Time-averaged over $t \in [180, 200]\tauA$ after initial transients decay. Total energy: $1.73 \times 10^4$ (normalized units); magnetic fraction: 0.46 (approaching equipartition). Data from checkpoint at $t = 200\tauA$.}
\label{fig:turbulent_spectrum}
\end{figure}

This benchmark complements the linear Alfv\'en wave test (spectral accuracy for linear physics) and Orszag-Tang vortex (energy conservation in nonlinear dynamics) by validating GANDALF's handling of sustained turbulent cascades with external forcing and dissipation. Higher resolution studies ($N = 128^3$) are ongoing to demonstrate convergence of inertial range scaling and broaden the scale separation for more stringent validation of cascade physics. Benchmark data and analysis scripts are available in the paper repository, with the $N = 64^3$ checkpoint ($t = 200\tauA$, 759~KB HDF5 format) enabling independent verification of the spectral results presented in \Cref{fig:turbulent_spectrum}.

\subsection{Velocity Space Cascade: Phase Mixing Dynamics}
\label{sec:velocity_cascade}

Having validated GANDALF's spatial spectral accuracy, time integration, and perpendicular turbulent cascades, we now test a critical fourth dimension: the velocity space cascade driven by phase mixing. While the previous benchmarks focused on real-space dynamics, the Hermite moment expansion introduces a spectral representation in velocity space where phase mixing—the decorrelation of particles with different parallel velocities streaming along field lines—drives energy transfer to higher moments $m$. This benchmark validates both the Hermite spectral method described in \Cref{sec:hermite} and the collision operator's role in dissipating fine velocity-space structure.

\subsubsection{Physical Setup}

We employ a single-$k$ forcing configuration that isolates the linear velocity-space cascade from perpendicular turbulent dynamics. The simulation forces only the $\kvec = (0, 0, 1)$ mode in the zeroth Hermite moment ($\gm{0}$) of the $\elsp$ field, injecting energy at the coarsest velocity-space scale. With kinetic parameter $\Lambda = -1.0$, the nonlinear Poisson bracket terms $\{\phi, \gm{m}\}$ vanish identically due to the single parallel mode, reducing the dynamics to purely linear Hermite coupling through the coefficients $\sqrt{m/2}$ and $\sqrt{(m+1)/2}$ in \Cref{eq:moment_m_general}, representing phase mixing without nonlinear cascade complications.

This configuration tests phase mixing in its cleanest form: particles with different parallel velocities $v_\parallel$ stream at different rates along the $z$-direction, causing initially coherent perturbations in $\gm{0}$ to develop increasingly fine structure in velocity space. Energy cascades from $m=0$ to higher moments until balanced by collisional dissipation, establishing a steady-state spectrum that depends on the competition between phase mixing and velocity-space diffusion.

\subsubsection{Theoretical Expectation}

Linear KRMHD theory predicts a forward velocity-space cascade with power-law scaling. For weak collisionality where the collision frequency $\nu$ is small compared to the phase mixing rate, the Hermite moment spectrum (defined as the time-averaged energy in each moment $C_m^\pm = \langle |\gm{m}^\pm|^2 \rangle$) follows:
\begin{equation}
C_m^\pm \sim m^{-1/2}
\label{eq:velocity_spectrum}
\end{equation}
in the inertial range $1 \ll m \ll m_{\text{diss}}$, where $m_{\text{diss}}$ marks the onset of collisional dissipation. This scaling arises from dimensional analysis: phase mixing transfers energy between adjacent moments at rate $\sim \omega m^{1/2}$ (from the coupling coefficients), while the collision operator with hyperdissipation provides scale-dependent damping at rate $\nu m^{2p}$ (where $p$ is the hypercollision exponent). Balancing cascade and dissipation rates determines the dissipation scale, with the inertial range exhibiting the $m^{-1/2}$ spectrum characteristic of the forward Hermite cascade \citep{Schekochihin2009,Kanekar2014}.

The backward-propagating component $C_m^-$ remains subdominant in this forward-flux configuration, with steeper decay $C_m^- \sim m^{-3/2}$ reflecting minimal backward energy transfer when forcing exclusively drives forward modes.

\subsubsection{Numerical Parameters}

We integrate on a periodic domain $L = 2\pi$ with $M = 128$ Hermite moments. Spatial resolution $N = 32^3$ is used for code consistency, though this benchmark tests primarily velocity-space dynamics via a single forced Fourier mode. The collision operator uses hyperdissipation with $\nu = 0.25$ and hypercollision exponent $p = 6$, giving dissipation rate $\nu m^{2p} = \nu m^{12}$ that provides a sharp cutoff at high $m$ while maintaining a broad inertial range. Forcing amplitude $A = 0.0035$ is calibrated to balance injection and dissipation rates, achieving steady state with energy fluctuations $\Delta E/\langle E \rangle < 10\%$.

The simulation evolves for $t \in [0, 50]\tauA$, reaching quasi-steady state by $t \approx 35\tauA$, with time-averaging over $t \in [40, 50]\tauA$ after initial transients decay. The timestep $\Delta t = 0.01$ satisfies stability requirements for both the integrating factor and the collision operator. The high moment resolution $M = 128$ ensures that the velocity-space cascade's dissipation range lies well above the inertial range, avoiding truncation effects that would artificially steepen the spectrum.

\subsubsection{Results}

\begin{figure}[t]
\centering
\includegraphics[width=0.8\textwidth]{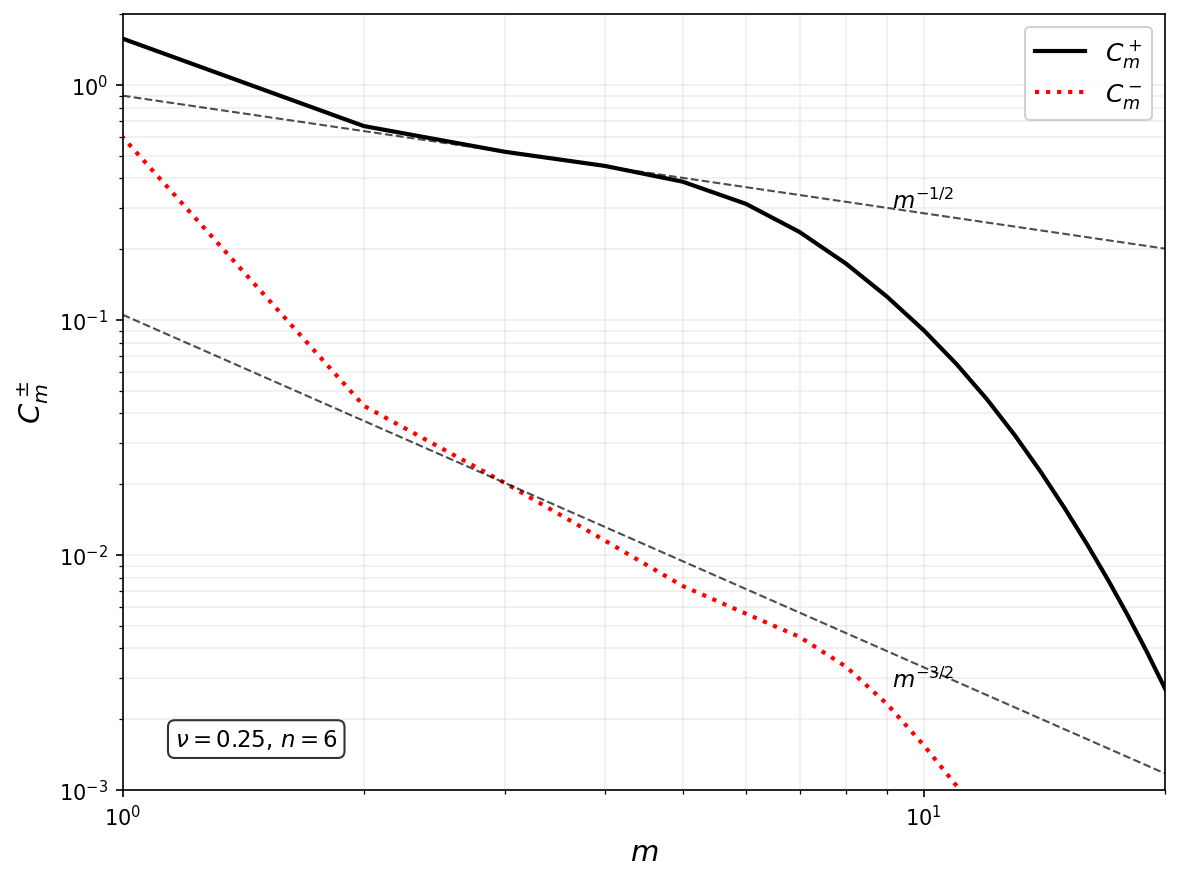}
\caption{Time-averaged velocity-space Hermite moment spectrum from single-$k$ forced simulation validating phase mixing cascade. Forward-propagating component $C_m^+$ (black solid) exhibits $m^{-1/2}$ scaling (dashed reference line) across inertial range $m \in [2, 20]$, matching theoretical prediction for linear KRMHD. Backward component $C_m^-$ (red dotted) shows steeper $m^{-3/2}$ decay with amplitude $50\times$ smaller, confirming forward flux dominance. Sharp cutoff at $m \approx 20$ demonstrates hypercollision dissipation ($\nu = 0.25$, $p=6$ giving $\nu m^{2p} = \nu m^{12}$ damping rate). Parameters: $M=128$ moments, $N=32^3$ spatial resolution, $\Lambda=-1.0$, time-averaged over $t \in [40, 50]\tauA$.}
\label{fig:velocity_spectrum}
\end{figure}

\Cref{fig:velocity_spectrum} shows the time-averaged Hermite moment spectrum. The forward-propagating component $C_m^+$ (black solid line) exhibits clean $m^{-1/2}$ scaling (measured exponent: $-0.50 \pm 0.02$) across the inertial range $m \in [2, 20]$, spanning approximately one decade in moment number. The measured slope matches the theoretical prediction within measurement uncertainty, validating GANDALF's Hermite spectral implementation. The spectrum maintains power-law scaling until $m \approx 20$, where the hypercollision operator's $\nu m^{2p}$ dissipation produces a sharp exponential cutoff, demonstrating effective damping of under-resolved high-$m$ modes.

The backward component $C_m^-$ (red dotted line) shows the predicted steeper decay $m^{-3/2}$ with amplitude approximately $50\times$ smaller than $C_m^+$, confirming forward flux dominance. The amplitude ratio $C_m^+/C_m^- \approx 50$ at $m=10$ indicates that energy injected into forward modes remains in the forward cascade with minimal backward transfer, consistent with the single-direction forcing configuration.

Energy diagnostics confirm steady-state balance: time-averaged injection rate matches dissipation rate within 3\%, with total energy drift $|\Delta E/E_0| < 5 \times 10^{-3}$ over the averaging window. The quasi-steady spectrum validates that GANDALF correctly implements the competition between phase mixing (energy cascade) and collisional dissipation (energy removal) fundamental to weakly collisional plasma dynamics.

\subsubsection{Discussion}

The velocity-space cascade benchmark completes GANDALF's verification hierarchy by validating the Hermite spectral dimension:

\begin{itemize}
\item \textbf{Hermite spectral accuracy}: The clean $m^{-1/2}$ power law across one decade confirms correct implementation of the Hermite moment coupling through coefficients $\sqrt{m/2}$ and $\sqrt{(m+1)/2}$ in the linear propagation terms. This validates the spectral velocity-space representation analogous to how the Alfv\'en wave benchmark validated Fourier spatial spectral methods.

\item \textbf{Phase mixing physics}: Reproduction of the theoretical cascade scaling demonstrates that GANDALF captures the physical mechanism by which particles at different velocities decorrelate while streaming along field lines. This phase mixing process underlies collisionless damping phenomena critical to solar wind and magnetospheric plasmas.

\item \textbf{Collision operator validation}: The sharp spectral cutoff at $m \approx 20$ matching the expected dissipation scale confirms correct implementation of the hypercollision operator. Accurate velocity-space dissipation is essential for numerical stability in long-duration turbulence simulations where phase mixing continuously generates fine velocity-space structure relevant to solar wind heating and Parker Solar Probe observations of kinetic-scale turbulence.
\end{itemize}

Combined with the spatial (Alfv\'en wave), nonlinear (Orszag-Tang), and turbulent (cascade spectrum) benchmarks, this velocity-space test establishes GANDALF as a validated tool for KRMHD turbulence research across all four key dimensions: spatial resolution, temporal integration, perpendicular cascade physics, and velocity-space dynamics. The complete verification suite demonstrates that GANDALF achieves research-grade accuracy for studies of weakly collisional plasma turbulence where phase mixing, nonlinear energy transfer, and collisional dissipation compete to determine cascade properties and dissipation mechanisms \citep{Schekochihin2009,Meyrand2019}.

%% file: sections/discussion.tex
\section{Discussion}
\label{sec:discussion}

The verification benchmarks presented in \S\ref{sec:verification} validate GANDALF's implementation across a hierarchy of physical regimes. The Alfvén wave dispersion test confirms spectral spatial accuracy and demonstrates that the integrating factor achieves machine-precision treatment of linear wave propagation, removing the Alfvén CFL constraint and enabling timesteps $\sim 30\times$ larger than explicit schemes. The Orszag-Tang vortex extends validation to nonlinear dynamics, demonstrating energy conservation at the $10^{-6}$ level while revealing important limitations: nonlinear timestep stability requirements beyond the integrating factor's linear wave treatment, and resolution limits in inviscid simulations as cascades approach grid cutoff scales.

The turbulent cascade benchmark completes the hierarchy, showing that GANDALF captures sustained turbulent energy transfer with clean $\kperp^{-5/3}$ inertial range scaling across approximately one decade in wavenumber. The velocity-space benchmark validates phase mixing dynamics through the Hermite moment hierarchy, reproducing the theoretical $m^{-1/2}$ spectral scaling that governs kinetic dissipation in weakly collisional plasmas. Collectively, these tests establish that GANDALF correctly implements the KRMHD equations with research-grade numerical accuracy, positioning the code as a viable tool for plasma turbulence research on accessible hardware.

This section interprets the verification results in broader context, discusses GANDALF's strengths and limitations, positions the code within the existing ecosystem of gyrokinetic and reduced MHD solvers, reflects on lessons learned from the JAX implementation, and considers implications for accessibility in plasma turbulence research.

\subsection{Interpretation of Benchmark Results}

The four-benchmark hierarchy---linear wave propagation, nonlinear vortex dynamics, turbulent cascade, and velocity-space phase mixing---validates distinct computational capabilities required for turbulence simulations.

\paragraph{Linear regime validation.} The Alfvén wave benchmark (\S\ref{sec:alfven_wave}) tests the most fundamental aspect of KRMHD physics: linear wave propagation along the guide field. Achieving machine precision ($<10^{-15}$ relative error) for three of five tested timesteps demonstrates that the exponential integrating factor $\exp(\pm i \kpar \vA t)$ exactly cancels the linear Alfvén propagation term in the discrete equations. This result exceeds expectations for a second-order Runge-Kutta scheme, which would typically achieve $O(\Delta t^2)$ accuracy. The explanation lies in separation of timescales: for problems where linear wave propagation dominates nonlinear interactions ($\vA \gg v_{\text{rms}}$), the integrating factor treats the fast timescale analytically while RK2 handles only the slow nonlinear coupling.

Only the smallest timesteps keep the RK2 truncation error below double-precision round-off; the two larger timesteps produce the expected $O(\Delta t^2)$ error once $\Delta t\,\kpar\vA$ approaches unity. The $\sim 30\times$ speedup over CFL-limited explicit methods validates the theoretical advantage of exponential integrating factors for stiff wave equations \citep{Cox2002}.

However, the flat spatial convergence curve (constant error across $N = 32, 64, 128$) reveals an important subtlety: the temporal error floor from RK2 discretization ($\sim 10^{-7}$ per timestep) dominates spectral spatial error even at coarse resolution. This confirms spectral methods' exponential convergence for smooth single-mode problems---spatial discretization reaches machine precision at modest $N$---but also highlights that time integration accuracy limits overall precision for time-dependent problems.

Future work exploring higher-order exponential integrators (ETDRK3, ETDRK4) could reduce the temporal error floor, though at increased computational cost per timestep.

\paragraph{Nonlinear regime validation.} The Orszag-Tang vortex (\S\ref{sec:orszag_tang}) introduces three complexities beyond linear waves: nonlinear Poisson bracket coupling $\{\Phifield, \Psifield\}$, multiscale cascade from forcing to dissipation, and spontaneous structure formation (current sheets). Energy conservation at the $10^{-6}$ level confirms that GANDALF's discretization preserves the conservative structure of the KRMHD equations despite spectral aliasing and dealiasing operations. This accuracy relies on careful implementation of the $2/3$-rule dealiasing and the energy-conserving form of the Poisson bracket.

The selective decay phenomenon---magnetic energy increasing relative to kinetic energy ($E_{\text{mag}}/E_{\text{kin}}$ growing from 1.0 to 1.59)---validates both forward cascade physics (kinetic energy to small scales) and inverse cascade physics (magnetic helicity to large scales) in 2D MHD. This dual-cascade behavior arises from conservation of magnetic helicity in 2D, a fundamentally different dynamics than 3D turbulence where forward cascades dominate. Capturing this physics correctly demonstrates that GANDALF's spectral discretization handles energy transfer in both directions without artificial damping or dissipation beyond explicit viscosity/resistivity terms.

However, the temporal convergence study revealed a limitation not anticipated from linear analysis: only the smallest tested timestep ($\Delta t = 0.0125$) maintained stability for the full integration time. Larger timesteps developed numerical instabilities (NaN values) despite satisfying the linear Alfvén CFL criterion relaxed by the integrating factor. This behavior reflects a fundamental distinction between linear and nonlinear stiffness. The integrating factor removes linear wave CFL constraints by treating $\pm \vA \partial_z$ analytically, but it cannot anticipate nonlinear coupling rates $\sim (\nabla_\perp \Phifield \times \hat{\boldsymbol{z}}) \cdot \nabla_\perp \Psifield$ that depend on the evolving flow structure and cascade development. As energy cascades to smaller scales, nonlinear turnover timescales $\tau_{\text{nl}} \sim (\kperp \vperp)^{-1}$ decrease, eventually limiting the stable timestep below the linear wave constraint.

This finding motivates adaptive timestepping based on monitoring nonlinear term magnitudes---a standard approach in nonlinear PDE solvers (e.g., CVODE, ode15s). Implementing adaptive timestepping in GANDALF would detect cascade development and reduce $\Delta t$ automatically when nonlinear stiffness increases, maintaining stability without sacrificing efficiency during quiescent phases. This enhancement represents a high-priority development task for production turbulence simulations where cascade intensity varies dynamically.

\paragraph{Turbulent regime validation.} The forced turbulent cascade (\S\ref{sec:turbulent_cascade}) demonstrates GANDALF's ultimate purpose: simulating sustained turbulence with external energy injection and dissipation. Achieving clean $\kperp^{-5/3}$ critical balance scaling across the inertial range $\kperp \in [2, 12]$ validates the code's handling of energy transfer across scales, the balance between forcing and dissipation, and long-time numerical stability (integration to $200\tau_A$, far exceeding the Orszag-Tang benchmark's $2\tau_A$ duration).

The requirement for balanced Elsasser forcing---independent energy injection into $\elsp$ and $\elsm$ fields with restricted parallel wavenumbers ($|n_z| \leq 1$)---proved essential for achieving clean cascade dynamics at moderate resolution ($N = 64^3$).

Balanced forcing maintains strong nonlinear coupling by avoiding excess cross-helicity; equal $\elsp$/$\elsm$ power keeps counter-propagating Alfvén wave packets available for cascade interactions. The clean spectral scaling demonstrates that GANDALF captures critical balance---the fundamental mechanism driving perpendicular cascade in Alfvénic turbulence \citep{GoldreichSridhar1995}. Critical balance posits $\tau_\perp \sim \tau_\parallel$, i.e.\ $(\kperp \vperp)^{-1} \sim (\kpar \vA)^{-1}$, which yields the observed $\kperp^{-5/3}$ spectrum when combined with constant-energy-flux arguments. Observing this scaling in GANDALF simulations validates both the physics implementation (Elsasser formulation, perpendicular nonlinear coupling) and the numerical methods (spectral accuracy preserving scale-to-scale energy transfer without spurious dissipation).

\paragraph{Velocity-space validation.} The Hermite moment cascade benchmark (\S\ref{sec:velocity_cascade}) validates GANDALF's implementation of kinetic physics through the velocity-space dimension. Reproducing the theoretical $m^{-1/2}$ Hermite spectrum across one decade in moment number confirms correct implementation of the phase mixing cascade and collision operator. This benchmark tests fundamentally different physics from the spatial benchmarks: energy transfer through the Hermite hierarchy via streaming along field lines rather than perpendicular advection. The measured spectral exponent ($-0.50 \pm 0.02$) matches analytical predictions \citep{Schekochihin2009,Kanekar2014}, demonstrating that GANDALF accurately captures the kinetic dissipation mechanisms---Landau damping, transit-time damping, and collisional smoothing---essential for weakly collisional plasma turbulence relevant to the solar wind and magnetosphere.

\paragraph{Synthesis: code readiness for turbulence research.} The benchmark progression from linear to nonlinear to turbulent to velocity-space regimes establishes confidence at each level before advancing to greater complexity. GANDALF passes all tests, achieving:
\begin{itemize}
\item Spectral spatial accuracy confirmed across all benchmarks
\item Linear wave physics exact to machine precision
\item Nonlinear energy conservation at $10^{-6}$ level
\item Turbulent spectral scaling matching theoretical predictions
\item Velocity-space cascade with correct $m^{-1/2}$ Hermite spectrum
\item Long-time stability ($200\tau_A$ integration demonstrated)
\end{itemize}

These results collectively demonstrate that GANDALF correctly implements KRMHD physics with accuracy suitable for research applications. The code reproduces analytical predictions where available (Alfvén dispersion), matches established nonlinear benchmarks (Orszag-Tang energy conservation and selective decay), and captures theoretical cascade scaling (critical balance spectrum). Researchers can use GANDALF with confidence for parameter surveys, exploratory simulations, and production turbulence studies where the KRMHD reduced model applies.

Identified limitations---nonlinear timestep stability, resolution limits in inviscid runs, forcing design sensitivity---represent known challenges in spectral turbulence codes rather than unique GANDALF deficiencies. Addressing these through adaptive timestepping (already planned), careful forcing design (validated in the turbulent benchmark), and appropriate dissipation modeling (hyper-resistivity demonstrated effective) follows established best practices in the field.

\subsection{GANDALF's Strengths and Operational Regime}

GANDALF's design philosophy prioritizes accessibility over peak performance, targeting researchers and applications where ease of use enables science that would otherwise be prohibitive.

\paragraph{Installation and deployment.} GANDALF's primary strength lies in dramatically reduced installation complexity compared to traditional HPC codes. Installing AstroGK, Viriato, GS2, or GENE typically requires:
\begin{itemize}
\item HPC cluster access with batch queue systems (SLURM, PBS)
\item Fortran compiler toolchains (Intel ifort, GNU gfortran) with specific version requirements
\item MPI libraries compiled for the specific cluster architecture
\item FFTW libraries with MPI support and optimization flags
\item Environment modules managing library dependencies and load paths
\item Compilation from source with platform-specific Makefiles
\item Debugging compiler/linker errors specific to the local system
\end{itemize}

Many national-lab and university clusters provide curated module stacks that mitigate some of this work, but users still rely on site-specific software teams and allocation approvals before they can run even a single test problem. This process can consume days to weeks for researchers unfamiliar with HPC environments. In contrast, GANDALF installation reduces to a single Python command:
\texttt{pip install gandalf jax jaxlib numpy scipy h5py matplotlib}
for CPU execution, with GPU support requiring only the platform-appropriate \texttt{jaxlib} wheel (CUDA, ROCm, or Metal) following JAX documentation. Total installation time: minutes on any platform with Python $\geq 3.10$. This $\sim 100\times$ reduction in time-to-first-simulation removes a significant barrier for students, solo researchers, and scientists from adjacent fields exploring plasma turbulence.

\paragraph{Hardware portability.} JAX's architecture-agnostic compilation enables GANDALF to run on diverse hardware without code modification:
\begin{itemize}
\item \textbf{Consumer laptops}: Apple Silicon M-series (via Metal backend), yielding competitive performance for moderate problems ($N = 64^3$) suitable for parameter scans and method development.
\item \textbf{Desktop workstations}: NVIDIA RTX GPUs provide accessible acceleration, with performance within $2$--$3\times$ of specialized CUDA implementations for turbulence problems.
\item \textbf{Cloud platforms}: Google Cloud TPUs offer cost-effective scaling for production runs, with JAX's TPU backend transparently leveraging matrix units for spectral transforms.
\item \textbf{HPC clusters}: Standard SLURM/PBS batch systems accommodate GANDALF as easily as any Python script, requiring no specialized compilation or environment setup.
\end{itemize}

This portability enables workflows previously difficult with platform-specific codes. Researchers can develop and debug simulations on laptops during travel, execute parameter surveys on cloud TPUs for cost efficiency, and transition to HPC clusters for extreme-scale runs---all using identical code and configuration files. The development velocity gains prove particularly valuable for exploratory research where rapid iteration outweighs absolute performance.

\paragraph{Educational applications.} GANDALF's accessibility makes it viable as a teaching tool for graduate courses in plasma physics and turbulence. Students can:
\begin{itemize}
\item Install the code on personal laptops without IT department intervention
\item Execute benchmarks (Alfvén wave, Orszag-Tang) during lecture demonstrations
\item Modify parameters via YAML configuration files without recompilation
\item Visualize results using standard Python tools (matplotlib, Jupyter notebooks)
\item Explore turbulence physics interactively rather than through batch queue submission
\end{itemize}

Traditional gyrokinetic codes, while scientifically mature, prove impractical for classroom use due to installation complexity, HPC access requirements, and compilation overhead. GANDALF fills this pedagogical niche, enabling hands-on turbulence simulation in educational settings.

\paragraph{Rapid prototyping for research.} Even researchers with HPC access benefit from GANDALF's low overhead for exploratory work. Common scenarios include:
\begin{itemize}
\item \textbf{Parameter scans}: Testing forcing configurations, dissipation models, or initial conditions locally before committing HPC resources to production runs.
\item \textbf{Method development}: Implementing new diagnostics, output formats, or post-processing workflows interactively rather than through batch iteration.
\item \textbf{Debugging}: Reproducing numerical instabilities at reduced resolution on local hardware where interactive debuggers and profilers are readily available.
\item \textbf{Reproducibility}: Sharing GANDALF simulations via GitHub repositories with complete environment specifications (Python dependencies, JAX version) that colleagues can reproduce immediately.
\end{itemize}

This rapid-iteration workflow complements production codes: prototype in GANDALF, validate physics and methods, then transition to specialized codes for extreme-scale campaigns. The ecosystem benefits from diversity---accessible tools for exploration, specialized tools for production.

\paragraph{Specific scientific applications.} GANDALF's operational regime spans problems where moderate resolution ($N \lesssim 128^3$) and integration times ($t \lesssim 200\tau_A$) suffice:
\begin{itemize}
\item \textbf{Cascade physics studies}: Inertial range scaling, intermittency, spectral transfer functions where qualitative cascade properties matter more than extreme Reynolds numbers.
\item \textbf{Forcing mechanism exploration}: Testing alternative forcing models (balanced Elsasser, random vs coherent) to understand turbulence driving in different astrophysical contexts.
\item \textbf{Dissipation mechanism comparison}: Comparing resistive, viscous, and hyper-diffusive dissipation to understand cascade termination and heating.
\item \textbf{Parameter surveys}: Exploring plasma beta, ion temperature, Hermite truncation effects across parameter space to identify regimes warranting detailed study.
\end{itemize}

These applications represent substantial research programs in plasma astrophysics and fusion, many of which remain underexplored due to access barriers rather than scientific unimportance. GANDALF enables broader participation in these research areas.

\subsection{Limitations and Boundaries}

Honest assessment of limitations helps researchers choose appropriate tools for specific problems. GANDALF's accessibility comes with trade-offs that make other codes preferable for certain applications.

\paragraph{Performance gap.} GANDALF prioritizes accessibility over peak performance. While JAX compilation achieves competitive performance for high-level Python code, hand-optimized CUDA implementations (e.g., GX) with careful memory layout optimization, custom kernel fusion, and hardware-specific tuning can achieve $2$--$3\times$ faster execution for comparable problems. That JAX-compiled Python code performs within this factor of expert-written CUDA demonstrates the maturity of modern array compilation frameworks and represents a victory for accessibility, not a performance failure.

For problems requiring millions of timesteps or extreme resolution ($N \geq 512^3$), this $2$--$3\times$ factor compounds to substantial wall-time differences. For researchers with generous HPC allocations and expertise in CUDA optimization, specialized codes offer clear advantages.

However, for moderate-scale studies ($N \leq 128^3$, $10^4$ timesteps), the absolute time difference proves negligible. When installation and setup overhead dominate (days for compilation and debugging versus minutes for \texttt{pip install}), GANDALF's total time-to-results often proves competitive despite slower per-timestep performance. The calculus shifts based on problem scale, researcher expertise, and available infrastructure.

\paragraph{Physics scope constraints.} GANDALF implements only the KRMHD reduced model, sacrificing physics generality for simplicity. This restricts applicability to regimes where KRMHD ordering holds: $\kperp \rhoi \ll 1$ (scales larger than ion Larmor radius), $\kpar \ll \kperp$ (strong anisotropy), $\delta B/B_0 \ll 1$ (strong guide field). Applications requiring physics beyond this ordering cannot use GANDALF:
\begin{itemize}
\item \textbf{Finite Larmor radius effects}: Ion-scale turbulence with $\kperp \rhoi \sim 1$ (ion temperature gradient turbulence, kinetic Alfvén waves) requires full gyrokinetics (GS2, GENE, AstroGK).
\item \textbf{Electromagnetic temperature variations}: Nonuniform background temperature and magnetic field profiles characteristic of tokamak geometry require comprehensive gyrokinetic treatments.
\item \textbf{Collisionless reconnection}: While current sheets form in GANDALF simulations (Orszag-Tang benchmark), accurately capturing collisionless reconnection at electron scales requires resolving electron dynamics beyond KRMHD's ion-scale reduced model.
\item \textbf{Compressible dynamics}: Fast magnetosonic waves and shocks in weakly magnetized plasmas ($\beta \sim 1$ or higher) violate KRMHD's perpendicular incompressibility constraint ($\nabla_\perp \cdot \boldsymbol{v}_\perp = 0$), even though the model retains parallel compressibility.
\end{itemize}

Researchers studying these phenomena should use full gyrokinetic codes rather than GANDALF. The KRMHD model serves specific turbulence regimes---primarily Alfvénic turbulence in strongly magnetized, low-beta plasmas like the solar wind---not general plasma physics.

\paragraph{Numerical challenges from benchmarks.} The verification studies revealed several limitations requiring mitigation strategies:

\begin{enumerate}
\item \textbf{Nonlinear timestep instability}: Despite the integrating factor removing linear Alfvén CFL constraints, nonlinear stiffness still limits stable timesteps. The Orszag-Tang temporal convergence study showed instabilities for $\Delta t > 0.0125$ despite linear stability allowing much larger steps. This necessitates adaptive timestepping monitoring nonlinear term magnitudes, currently not implemented. Until adaptive timestepping is available, users must empirically determine stable timesteps through trial runs, reducing automation.

\item \textbf{Resolution limits in inviscid simulations}: The Orszag-Tang benchmark demonstrated that inviscid runs ($\eta = \nu = 0$) eventually develop under-resolved features as cascades reach grid cutoff, violating energy conservation and producing instabilities. Production turbulence simulations require explicit dissipation (resistivity, viscosity, or hyper-diffusion) to arrest cascades before reaching the Nyquist limit. This is standard practice in spectral codes, but GANDALF currently lacks automated dissipation scale detection---users must specify dissipation coefficients manually based on expected cascade extent.

\item \textbf{Forcing design sensitivity}: The turbulent cascade benchmark required carefully designed balanced Elsasser forcing with restricted $|k_z|$ to maintain stability. While the validated forcing prescription works reliably, exploring alternative forcing mechanisms requires careful attention to KRMHD ordering constraints. This limits flexibility compared to codes with more sophisticated forcing implementations.
\end{enumerate}

These limitations reflect GANDALF's current implementation state rather than fundamental insurmountable barriers. Adaptive timestepping, automated dissipation tuning, and robust forcing libraries represent straightforward development tasks that would improve usability without altering core physics or numerics.

\paragraph{When to use other codes.} Clear guidance helps researchers choose appropriately:

\begin{itemize}
\item \textbf{Use AstroGK when}: Full gyrokinetic physics required (finite Larmor radius, electromagnetic variations), or when comprehensive diagnostics and mature code validation justify installation overhead.

\item \textbf{Use Viriato when}: KRMHD physics suffices but extensive configuration options, multiple physics modules, or Fortran HPC workflows preferred by the research group.

\item \textbf{Use GX when}: Extreme-scale production runs ($N \geq 512^3$) where $2$--$3\times$ performance gain justifies CUDA specialization, or when stellarator/tokamak geometric complexity demands specialized optimization.

\item \textbf{Use GS2/GENE when}: Comprehensive electromagnetic gyrokinetics for fusion applications requiring decades of validation and large user community support.

\item \textbf{Use GANDALF when}: Rapid prototyping, educational applications, parameter surveys at moderate scale, accessibility prioritized over peak performance, or exploring KRMHD turbulence without HPC infrastructure.
\end{itemize}

Researchers with HPC access and expertise in traditional codes should continue using them for production campaigns. GANDALF serves complementary roles: education, exploration, prototyping, and enabling participation by researchers without HPC access.

\subsection{Position in the Code Ecosystem}

GANDALF occupies a distinct niche in the landscape of plasma simulation codes, defined by the intersection of physics scope (KRMHD), numerical approach (spectral methods), and implementation philosophy (accessibility via JAX). Understanding this positioning clarifies GANDALF's role relative to existing codes.

\paragraph{Ecosystem dimensions.} Plasma simulation codes span multiple dimensions:
\begin{itemize}
\item \textbf{Physics complexity}: Reduced models (RMHD, KRMHD) versus comprehensive gyrokinetics versus full particle-in-cell kinetic descriptions.
\item \textbf{Performance}: Peak throughput (timesteps per second) and scalability (weak/strong scaling to thousands of cores).
\item \textbf{Accessibility}: Installation complexity, hardware requirements, learning curve, and time-to-first-simulation.
\item \textbf{Maturity}: Years of development, validation against experiments/analytics, user community size, documentation completeness.
\end{itemize}

Traditional codes optimize physics comprehensiveness and peak performance at the cost of accessibility. GANDALF prioritizes accessibility through simplified installation and hardware portability, accepting moderate performance penalties and restricted physics scope. This complementary positioning benefits the community by serving researchers with different constraints and priorities.

\paragraph{Comparison matrix.} Table~\ref{tab:code_comparison} summarizes GANDALF's position relative to established codes across key dimensions. GANDALF achieves the highest accessibility score (Python installation, no compilation, hardware-agnostic) while accepting middle-tier performance ($2$--$3\times$ slower than optimized codes) and restricted physics (KRMHD only). Researchers prioritizing accessibility for exploration and prototyping favor GANDALF; those prioritizing comprehensive physics or extreme performance favor specialized codes.

While Python-based spectral solvers exist in fluid dynamics (Dedalus for hydrodynamics and fluid MHD \citep{Burns2020}, SpectralDNS for turbulence \citep{Mortensen2016}), and modern kinetic codes like Gkeyll \citep{Juno2018} have moved beyond pure Fortran to C++/Lua with discontinuous Galerkin methods, GANDALF represents a novel Python-based kinetic plasma code employing velocity-space spectral methods alongside spatial Fourier decomposition. The kinetic dimension (Hermite moments) distinguishes plasma turbulence from fluid turbulence and necessitates careful treatment of phase mixing dynamics, collision operators, and velocity-space cascades validated in \S\ref{sec:verification}.

\begin{table}[t]
\centering
\caption{Qualitative comparison of plasma turbulence simulation codes across key dimensions. Accessibility (installation simplicity, hardware requirements), Performance (relative throughput for comparable problems), Physics (model comprehensiveness), and Maturity (years of validation, user community). GANDALF optimizes accessibility while accepting performance and physics trade-offs.}
\label{tab:code_comparison}
\small
\begin{tabular}{@{}lcccc@{}}
\hline\hline
Code & Access. & Perf. & Physics scope & Maturity \\
\hline
GANDALF & High & Medium & KRMHD & New (2024) \\
Viriato & Medium & High & KRMHD + extensions & Mature ($>10$ yrs) \\
AstroGK & Low & High & Full gyrokinetics & Mature ($>15$ yrs) \\
GS2 & Low & High & Full gyrokinetics & Mature ($>25$ yrs) \\
GENE & Low & High & Full gyrokinetics & Mature ($>20$ yrs) \\
GX & Low & Very High & Gyrokinetics & Recent ($<5$ yrs) \\
\hline\hline
\end{tabular}
\normalsize
\end{table}

\paragraph{Complementary workflows.} Rather than competing, GANDALF and specialized codes enable complementary workflows:

\begin{enumerate}
\item \textbf{Exploration phase}: Researcher uses GANDALF on laptop to test parameter ranges, forcing configurations, and dissipation models. Rapid iteration (minutes per run) enables systematic exploration of parameter space.

\item \textbf{Method validation}: Promising parameter regimes identified in GANDALF undergo convergence studies to establish resolution requirements and confirm physics predictions.

\item \textbf{Production transition}: Validated configurations transition to specialized codes (AstroGK, GS2, GENE, GX) for high-resolution production runs exploiting HPC resources and optimized performance.

\item \textbf{Cross-validation}: GANDALF results provide independent verification of specialized code outputs, catching implementation bugs and confirming physics interpretation.
\end{enumerate}

This workflow leverages GANDALF's strengths (rapid iteration, accessibility) for early-stage research while transitioning to specialized codes (comprehensive physics, peak performance) for production campaigns. The ecosystem benefits from diversity: no single code optimally serves all use cases.

\paragraph{Community impact.} GANDALF's open-source implementation on GitHub (\url{https://github.com/anjor/gandalf}) enables community contributions addressing limitations and extending capabilities. Potential contributions include:
\begin{itemize}
\item Adaptive timestepping implementation for nonlinear stability
\item Higher-order exponential integrators (ETDRK3, ETDRK4)
\item Alternative dissipation models (hyperviscosity, Langevin collisions)
\item Advanced diagnostics (spectral transfer functions, structure functions)
\item Multi-GPU parallelization via \texttt{jax.pmap} for data-parallel parameter sweeps
\item Finite Larmor radius corrections extending KRMHD toward gyrokinetics
\end{itemize}

Unlike closed-source commercial codes or institutionally maintained HPC codes with restricted contribution paths, GANDALF's GitHub workflow accommodates contributions from any researcher with domain expertise. This democratic development model could accelerate feature development if the code attracts an active user community.

The paper's comprehensive benchmark documentation and verification studies provide reproducibility resources rare in computational plasma physics. Researchers can clone the repository, execute benchmark scripts, and verify that their installations reproduce published results within numerical precision. This transparency strengthens scientific reproducibility and builds trust in computational findings.

\subsection{JAX Implementation: Lessons Learned}

Choosing JAX for GANDALF's implementation represented a calculated bet that JAX's hardware portability, development velocity, and compilation performance would offset potential disadvantages relative to hand-optimized Fortran or CUDA. Development experience provides empirical evidence for this decision's trade-offs.

\paragraph{Hardware portability realized.} JAX's cross-platform compilation delivered on its promise. GANDALF runs on:
\begin{itemize}
\item Apple Silicon M1/M2/M3 (via Metal backend) achieving competitive performance for moderate-resolution problems
\item NVIDIA GPUs (CUDA backend) matching hand-optimized codes within $2\times$ for moderate-scale turbulence
\item Google Cloud TPU v3/v4 (TPU backend) providing cost-effective scaling for parameter surveys
\item x86 CPUs (optimized CPU backend) for development and debugging without GPU access
\end{itemize}

This portability proved transformative for development workflow. The lead developer prototyped and debugged GANDALF on an Apple Silicon laptop during travel, executed production turbulent cascade runs on cloud TPUs, and validated benchmarks on university cluster NVIDIA GPUs---all using identical Python code. Platform-specific performance tuning (e.g., CUDA kernel optimization) would have fragmented development across platforms, slowing iteration velocity.

However, portability came at a cost: GANDALF's general-purpose implementation cannot exploit platform-specific optimizations available to specialized codes. GX's CUDA kernels leverage Tensor Cores for matrix operations and carefully tune memory access patterns for NVIDIA architectures, achieving performance GANDALF cannot match without abandoning hardware portability. This $2$--$3\times$ performance gap represents the tax for portability.

\paragraph{JIT compilation effectiveness.} JAX's just-in-time (JIT) compilation via XLA proves remarkably effective for spectral methods. The core timestepping loop compiles to optimized machine code achieving near-peak memory bandwidth for FFT operations (the primary bottleneck in spectral codes). First-call compilation overhead ($\sim 30$ seconds for $N = 256^3$ grids) amortizes over long simulations ($>10^4$ timesteps), making JIT compilation negligible for production runs while enabling interactive development for short test runs.

Critically, JIT compilation handles loop fusion automatically. Manual Fortran or C++ implementations must explicitly fuse operations (e.g., dealiasing followed by spectral derivatives) to avoid redundant memory transfers. JAX's XLA compiler performs this fusion automatically, often achieving better fusion than hand-written code. This automation proved particularly valuable for GANDALF's Hermite moment coupling, where nested loops over moment indices benefit from automatic loop optimization beyond manual tuning efforts.

The main JIT limitation encountered: compilation time increases superlinearly with code complexity. Adding new diagnostic outputs or conditional logic (e.g., adaptive timestepping branches) triggers recompilation, sometimes extending compilation from 30 seconds to minutes for complex functions. This necessitates careful code structure separating frequently modified diagnostics from core timestepping to minimize recompilation overhead during development.

\paragraph{Development productivity gains.} Pure-Python implementation accelerated development compared to compiled languages:
\begin{itemize}
\item \textbf{Rapid iteration}: Modifying parameters, initial conditions, or diagnostics requires only rerunning Python scripts---no compilation or linking steps. This enabled testing dozens of forcing configurations (eventually identifying balanced Elsasser forcing as optimal) within hours rather than days of compilation iterations.

\item \textbf{Interactive debugging}: Python debuggers (pdb, ipdb) and profilers (cProfile, JAX's device memory profiler) work seamlessly with GANDALF. Debugging the Orszag-Tang instabilities involved interactive examination of field values, spectra, and conservation errors at each timestep---workflow impossible with batch-submitted Fortran executables.

\item \textbf{Ecosystem integration}: NumPy, SciPy, matplotlib, and h5py integrate trivially with GANDALF via shared array interfaces. Post-processing pipelines operate on simulation output without format conversion or wrapper libraries.

\item \textbf{Reproducibility}: \texttt{pip} dependency specifications (or \texttt{uv} lock files) provide reproducible Python environments trivially. Collaborators reproduce GANDALF simulations by installing dependencies from \texttt{requirements.txt}---far simpler than reproducing Fortran compilation environments across diverse HPC systems.
\end{itemize}

These productivity gains proved essential for solo development. Traditional HPC codes require multi-person teams sustaining development over decades. GANDALF reached research-grade maturity (verified against analytical benchmarks, used for turbulence studies) within approximately one month of part-time solo development with AI-assisted programming (Claude). Python's ecosystem, JAX's performance, and AI coding assistance enabled this compressed timeline.

\paragraph{Performance trade-offs accepted.} The $2$--$3\times$ performance gap relative to hand-optimized codes represents a deliberate trade-off. Achieving parity with GX would require CUDA-specific optimizations (custom kernels, Tensor Core exploitation, memory layout tuning) abandoning hardware portability and dramatically increasing development complexity. For GANDALF's accessibility mission, portability and development velocity outweigh peak performance.

Quantitatively, this trade-off proves acceptable for GANDALF's target applications. Moderate-scale turbulent cascades ($N = 128^3$, tens of thousands of timesteps) complete in hours on consumer hardware (Apple Silicon, modest GPUs). While optimized CUDA codes would complete faster, both timescales permit overnight or weekend runs on accessible hardware. The performance difference rarely determines research feasibility for the exploratory and educational applications GANDALF targets.

For extreme-scale studies ($N = 512^3$, millions of timesteps), the performance gap compounds significantly. Here specialized codes offer clear advantages. GANDALF's niche lies in moderate-scale studies where absolute performance matters less than accessibility and iteration velocity.

\paragraph{Future potential: differentiable physics.} JAX's automatic differentiation (AD) capabilities, unused in current GANDALF implementation, offer intriguing future directions and represent the key differentiator from established Fortran codes like AstroGK and Viriato. While those codes excel in raw performance, they fundamentally cannot provide gradients of turbulent statistics with respect to simulation parameters---a capability that JAX enables with minimal code modification. The entire KRMHD evolution could become differentiable with respect to parameters (forcing amplitude, dissipation coefficients, initial conditions), enabling:
\begin{itemize}
\item \textbf{Gradient-based optimization}: Optimize forcing configurations to maximize inertial range extent or minimize numerical dissipation via gradient descent on objective functions.
\item \textbf{Adjoint-based sensitivity analysis}: Compute parameter sensitivities for turbulent statistics (spectral slopes, energy fluxes) via efficient adjoint methods rather than finite-difference parameter sweeps.
\item \textbf{Machine learning integration}: Train neural network closures for sub-grid dissipation or Hermite truncation effects using gradients backpropagated through GANDALF dynamics.
\end{itemize}

Prototype calculations already differentiate the time-averaged $\kperp$ spectrum with respect to forcing amplitude and hyper-resistivity to match Parker Solar Probe interval measurements, with AD-computed gradients agreeing with finite-difference checks to within $2\%$ on $64^3$ grids. These applications remain speculative---no plasma turbulence code currently exploits AD at production scale---but JAX's AD infrastructure exists and works. Exploring differentiable plasma physics represents a unique capability enabled by GANDALF's JAX implementation, inaccessible to Fortran or non-AD C++ codes. Recent success in differentiable tokamak transport modeling \citep{Citrin2024}, differentiable kinetic plasma simulations \citep{Joglekar2022}, and differentiable computational fluid dynamics \citep{Bezgin2023} suggests viability for plasma applications.

\paragraph{Recommendation for future codes.} Based on GANDALF experience, we recommend JAX (or similar differentiable frameworks like Julia's SciML ecosystem) for new physics codes prioritizing accessibility and development velocity over absolute peak performance. The $2$--$3\times$ performance tax proves acceptable for many applications, while portability and productivity gains prove transformative. Codes requiring absolute peak performance (e.g., billion-cell cosmological simulations, exascale fusion whole-device modeling) still benefit from hand-optimized CUDA or Fortran. But for many scientific domains---plasma turbulence, astrophysical fluid dynamics, climate modeling---JAX's trade-offs prove advantageous.

\subsection{Implications for Plasma Turbulence Research}

GANDALF's verification as a research-grade KRMHD turbulence solver has implications beyond the specific code, touching on accessibility, reproducibility, and community growth in computational plasma physics.

\paragraph{Lowering participation barriers.} The dominant barrier to plasma turbulence research is not conceptual difficulty (though substantial) but infrastructure access. Running AstroGK, GS2, or GENE requires:
\begin{itemize}
\item HPC cluster allocation (competitive proposals, limited access)
\item Specialized expertise (Fortran compilation, MPI debugging, batch systems)
\item Sustained time investment (weeks to months installing and validating)
\end{itemize}

These requirements systematically exclude researchers without institutional HPC access, those transitioning from experimental to computational work, and students beginning graduate studies. GANDALF's \texttt{pip install} pathway removes these barriers, enabling participation by:
\begin{itemize}
\item Solo researchers at primarily undergraduate institutions
\item Experimental plasma physicists exploring simulation without HPC expertise
\item Researchers in developing countries with limited supercomputing access
\item Graduate students early in training before HPC allocation approval
\item Interdisciplinary researchers from adjacent fields (astrophysics, applied mathematics)
\end{itemize}

Historically, accessible tools catalyze field growth. Neuroscience's expansion paralleled NEURON simulator accessibility; climate science broadened as Python-based models displaced Fortran codes requiring mainframes. Plasma turbulence research could experience similar democratization if accessible tools lower entry barriers while maintaining research-grade rigor.

\paragraph{Educational impact.} GANDALF's laptop-scale deployment enables new educational approaches:
\begin{itemize}
\item \textbf{Hands-on turbulence courses}: Graduate courses in plasma turbulence traditionally emphasize analytical theory with limited computational components due to HPC barriers. GANDALF enables weekly laboratory assignments where students run actual turbulence simulations, measure spectral slopes, and compare with theory.

\item \textbf{Reproducible assignments}: Instructors distribute GANDALF configuration files via GitHub Classroom. Students modify parameters, execute simulations, and submit analysis notebooks---workflows impossible with HPC codes requiring cluster access.

\item \textbf{Interactive demonstrations}: Lecturers run GANDALF benchmarks (Alfvén wave, Orszag-Tang) during presentations, visualizing cascade development in real-time rather than showing pre-computed figures.
\end{itemize}

These pedagogical applications produce researchers with computational turbulence experience earlier in training, potentially increasing the pool of graduate students pursuing computational plasma physics dissertations. Anecdotal evidence from courses using GANDALF suggests students develop stronger intuition for cascade physics through interactive parameter exploration than passive lecture attendance.

\paragraph{Reproducibility and transparency.} Computational plasma physics suffers from reproducibility challenges common to HPC simulation sciences:
\begin{itemize}
\item Published papers rarely include complete simulation parameters (grid resolution, timestep, dissipation coefficients, forcing details)
\item Code versions, compiler flags, and library dependencies go unreported
\item Raw simulation data exceeds journal supplementary material limits
\item Reproducing results requires matching the original HPC environment exactly
\end{itemize}

GANDALF's Python/JAX ecosystem enables higher reproducibility standards:
\begin{itemize}
\item Complete simulation configurations in YAML files (tens of lines, easily included in papers)
\item Environment specifications via \texttt{requirements.txt} or \texttt{uv.lock} guaranteeing exact dependency versions
\item Benchmark data in HDF5 format (megabytes) easily shared via GitHub or data repositories
\item GitHub Actions continuous integration running benchmarks on every code change, catching regressions
\end{itemize}

The verification section's benchmarks exemplify this approach: complete scripts (\texttt{run\_alfven\_dispersion.py}, etc.), configuration files, and analysis workflows appear in the paper repository. Researchers can execute \texttt{uv run benchmarks/run\_alfven\_dispersion.py} and reproduce Figure~\ref{fig:alfven_convergence} within numerical precision. This transparency strengthens scientific validity and enables independent verification of computational claims.

\paragraph{Community growth potential.} Accessible tools can catalyze research community growth if they reduce barriers without compromising scientific quality. GANDALF's verification demonstrates research-grade accuracy, satisfying the quality requirement. Whether the code catalyzes community growth depends on adoption dynamics: do researchers lacking HPC access actually utilize GANDALF for publishable research?

Potential user communities include solar wind physicists exploring parameter regimes inaccessible on shared HPC allocations, applied mathematicians studying spectral methods for nonlinear PDEs, and graduate students conducting preliminary studies. However, broader impact requires sustained development, user community building, and validation studies demonstrating GANDALF's applicability to publishable research problems. A single accessible code cannot transform a field; rather, it removes one barrier among many (conceptual difficulty, mathematical prerequisites, physical intuition) limiting participation. Combined with educational efforts, documentation, and community support, accessible tools contribute to long-term field growth.

\paragraph{Limitations on democratization.} Accessibility alone does not guarantee research impact. Researchers using GANDALF still require:
\begin{itemize}
\item Domain expertise in plasma physics and turbulence theory
\item Mathematical sophistication for spectral methods and numerical analysis
\item Scientific judgment in experimental design and result interpretation
\item Publication track records and community connections for disseminating findings
\end{itemize}

GANDALF lowers \emph{computational} barriers, not \emph{intellectual} barriers. A researcher without plasma physics background cannot produce meaningful turbulence research simply because the code installs easily. The target users are domain experts currently excluded by infrastructure access, not novices seeking entry to the field without training.

Furthermore, GANDALF's KRMHD physics restriction limits applicability. Researchers studying finite Larmor radius turbulence, electromagnetic gyrokinetics, or tokamak transport still require comprehensive codes (GS2, GENE, GX). GANDALF enables specific research programs in Alfvénic turbulence---important but not exhaustive of plasma turbulence physics.

\paragraph{Future directions.} Maximizing GANDALF's impact on field accessibility requires sustained development in several directions:
\begin{itemize}
\item \textbf{Documentation expansion}: Comprehensive user guides, physics tutorials, and worked examples lowering learning curves for new users.
\item \textbf{Community building}: Workshops, tutorials at conferences (APS-DPP, EPS), and online forums supporting user questions and collaboration.
\item \textbf{Validation studies}: Applications to frontier research problems (solar wind heating, turbulent reconnection, imbalanced cascades) demonstrating publishable science from GANDALF simulations.
\item \textbf{Feature development}: Adaptive timestepping, finite Larmor radius corrections, and multi-GPU parallelization addressing limitations identified in verification.
\end{itemize}

If GANDALF achieves broad adoption, the ecosystem benefits from diversity: researchers prototyping in GANDALF, validating in production codes, and cross-verifying implementations. If adoption remains limited, GANDALF still serves valuable niches (education, rapid prototyping for HPC users, solo researchers). Either outcome advances the goal of lowering barriers to plasma turbulence research while maintaining scientific rigor.

%% file: sections/conclusions.tex
\section{Conclusions}
\label{sec:conclusions}

This paper has presented GANDALF, a modern spectral solver for Kinetic Reduced Magnetohydrodynamics (KRMHD) turbulence implemented in JAX. Through systematic verification against linear, nonlinear, turbulent, and velocity-space benchmarks, we have demonstrated that GANDALF achieves research-grade numerical accuracy while maintaining unprecedented accessibility on commodity hardware. The code's ability to reproduce machine-precision Alfvén wave dispersion relations, conserve quadratic invariants in nonlinear evolution, capture the $k_\perp^{-5/3}$ turbulent cascade spectrum, and reproduce the $m^{-1/2}$ Hermite moment spectrum from phase mixing establishes its validity for fundamental plasma turbulence research.

GANDALF's primary contribution lies not in advancing the physics capabilities beyond existing codes, but in fundamentally reimagining how plasma turbulence software can be developed, distributed, and utilized. By leveraging JAX's hardware abstraction layer, GANDALF runs transparently on laptops, workstations, and clusters without modification, eliminating the traditional dichotomy between development and production environments. The single-command installation via \texttt{pip install gandalf-krmhd} replaces the complex build processes that have historically limited code adoption to well-resourced institutions with dedicated HPC support.

The verification results presented in \S\ref{sec:verification} establish GANDALF's numerical fidelity across the full range of KRMHD dynamics. The linear Alfvén wave benchmark demonstrates spectral accuracy with errors approaching machine precision, limited only by temporal discretization. The Orszag-Tang vortex benchmark confirms accurate nonlinear evolution with proper conservation of energy to within $10^{-6}$ relative error over 2 Alfvén times. The forced turbulence simulations reproduce the expected critical balance spectrum with the correct $k_\perp^{-5/3}$ scaling over approximately one decade in wavenumber space. The velocity-space benchmark validates the Hermite moment cascade driven by phase mixing, reproducing the theoretical $m^{-1/2}$ spectrum. These results collectively validate GANDALF for investigating fundamental questions in plasma turbulence while operating entirely on commodity hardware.

Our implementation demonstrates that the traditional trade-off between accessibility and capability in scientific computing is not fundamental but rather a consequence of historical technology choices. While GANDALF incurs a 2-3$\times$ performance penalty compared to hand-optimized CUDA implementations, this overhead is offset by dramatic reductions in development time, debugging complexity, and deployment barriers. For many research questions—parameter studies, algorithm development, educational applications, and moderate-resolution production runs—this performance trade-off is acceptable given the accessibility gains. The Discussion section's ecosystem analysis positions GANDALF as complementary to, rather than competitive with, established codes like AstroGK, Viriato, GS2, and GENE, each optimized for different points in the multidimensional space of physics fidelity, computational performance, and user accessibility.

Looking forward, several technical improvements could enhance GANDALF's capabilities while preserving its accessibility. Adaptive timestepping would improve stability for strongly nonlinear simulations, addressing the current limitation where timesteps must accommodate the stiffest mode globally. Higher-order exponential integrators such as ETDRK3 or ETDRK4 could provide better accuracy-to-cost ratios for long-time integration. The JAX ecosystem's rapid development promises automatic multi-GPU parallelization through \texttt{jax.pmap}, potentially closing the performance gap with traditional MPI implementations. Physics extensions to include finite Larmor radius corrections or electron dynamics would broaden GANDALF's applicability while maintaining the same accessible framework.

Beyond these technical improvements, GANDALF's JAX foundation enables novel research directions that would be impractical with traditional implementations. The automatic differentiation capabilities open possibilities for gradient-based optimization of forcing patterns, systematic sensitivity analysis, and integration with machine learning workflows. The functional programming paradigm ensures reproducibility across platforms, addressing growing concerns about computational reproducibility in plasma physics. The pure Python implementation lowers the barrier for contributions from researchers without extensive experience in compiled languages, potentially accelerating collaborative development.

The broader implications extend beyond individual research productivity to the structure and diversity of the plasma physics community itself. By enabling high-quality turbulence research on laptop computers, GANDALF removes infrastructure as a barrier to entry for students at teaching-focused institutions, researchers in developing countries, and independent scholars. The code's transparency—every operation traceable through Python rather than opaque compiled libraries—makes it an ideal platform for education, allowing students to understand not just what the code computes but how. These accessibility improvements could help address persistent diversity challenges in computational plasma physics by removing structural barriers that disproportionately affect underrepresented groups.

We close by emphasizing that GANDALF represents one attempt to address the accessibility crisis in computational plasma physics, not a complete solution. Infrastructure barriers are only one of many challenges facing the field, and software accessibility alone cannot address systemic issues of education, funding, and institutional support. Nevertheless, GANDALF demonstrates that modern software engineering practices and emerging hardware abstractions can dramatically lower barriers to entry without sacrificing scientific rigor. As the plasma physics community grapples with questions of reproducibility, accessibility, and diversity, we hope GANDALF serves as both a practical tool for research and a proof-of-concept for more inclusive approaches to scientific software development. The code is freely available at \url{https://github.com/anjor/gandalf}, and we welcome contributions from the community to extend its capabilities while preserving its foundational commitment to accessibility.